\RequirePackage{fix-cm}
\documentclass[smallextended]{svjour3} 

\smartqed  
\usepackage{graphicx}
\usepackage{cite}
\usepackage[english]{babel}
\usepackage{amsmath}
\usepackage{amsfonts}
\usepackage{amssymb}
\usepackage{subfig}
\usepackage{hyperref}
\usepackage{multirow}
\usepackage[usenames, dvipsnames]{color}
\usepackage{bbold}
\usepackage{bbm}
\usepackage{tabularx}
\usepackage{amsfonts}
\usepackage{mathtools}

\usepackage{hyperref}
\usepackage[english]{babel}
\hypersetup{
    colorlinks=true,
    linkcolor=ForestGreen,
    filecolor=magenta,      
    urlcolor=cyan,
    citecolor=blue
}
\usepackage{caption}
\titlerunning{Short title}
\usepackage{epstopdf}
\usepackage{lipsum}

  \usepackage[toc,page]{appendix}
\begin{document}

\title{Quantum communication with $SU(2)$ invariant separable $2\times N$ level systems
}

\titlerunning{Quantum communication with $SU(2)$ invariant separable $2\times N$ level systems}        

\author{Sooryansh Asthana$^{*}$
        \and
           Rajni Bala \and
           V. Ravishankar
}


\institute{ \at
              Department of Physics, Indian Institute of Technology Delhi, New Delhi-110016, India. \\
              \email{sooryansh.asthana@physics.iitd.ac.in}           
}

\date{Received: date / Accepted: date}

\maketitle
\begin{abstract}
Information is encoded in a qubit in the form of its Bloch vector. 
In this paper, we propose protocols for remote transfers of information in a known and an unknown qubit to qudits using $SU(2)$-- invariant $2 \times N$- level separable discordant states as quantum channels.  These states have been identified as separable equivalents of the two-qubit entangled Werner states in [Bharath \& Ravishankar,  \href{https://doi.org/10.1103/PhysRevA.89.062110}{Phys. Rev. A 89, 062110}]. Due to $SU(2) \times SU(2)$ invariance of these states, the remote qudit can be changed by performing appropriate measurements on the qubit.  We also propose a protocol for transferring information of a family of unknown qudits to  remote qudits using  $2 \times N$--level states as  channels. Finally, we propose a protocol for swapping of quantum discord from $2\times N$-- level systems to $N \times N$-- level systems. All the protocols proposed in this paper involve separable states as quantum channels.  
\end{abstract}

\maketitle

\section{Introduction}
Recent times have witnessed a surge of interest in the study of novel and distinctive features of different quantum correlations, e.g., quantum nonlocality  \cite{Bell64}, quantum entanglement \cite{Horodecki09}, steering \cite{Wiseman07}, and quantum discord \cite{Ollivier01}, etc. The interest owes to several nonclassical tasks in which they act as resources. Examples include quantum teleportation \cite{Bennett93}, superdense coding \cite{Bennett92}, remote state preparation \cite{Bennett01}, device-independent quantum cryptography \cite{Vazirani14} and entanglement swapping \cite{Zukowski93}.  In these protocols, either only entangled states are resourceful or they yield an advantage, manifested as a better performance of the protocol (e.g., teleportation fidelity \cite{Popescu94a}). 

Promising though these applications are, the challenges in the generation of entangled states and sustaining them for a long duration act as restrictions in these tasks \cite{Huang11, Monz11, Yao12}. 
 This has led to the study of nonclassical communication tasks that can be performed using nonclassical correlations beyond entanglement, {\it viz.}, quantum discord \cite{Dakic12, Madhok13} and geometric discord, etc. Yet, in all these approaches, there is a sharp decline in fidelity of the transmitted state with the input state, as compared to the scenario in which an entangled state is put to use \cite{Popescu94a, Fonseca19}.
 
 In a parallel approach, there have been many theoretical proposals and successful attempts to simulate (or perform) quantum computation \cite{Perez16}, quantum search algorithm \cite{Garcia18}, quantum information processing \cite{Spreeuw01}, quantum random walk \cite{Goyal13} with classical waves. The employment of classical waves for their implementation overcomes the problem of sustaining quantum coherence for a long duration. In spite of these advances, a major glitch is that quantum nonlocality cannot be simulated in the classical domain. It restricts the implementation of quantum communication protocols involving nonlocal states using classical waves.

These difficulties have led us to ask the question: do there exist {\it separable} resource states in $2 \times N$ dimensions, which can be used for transferring the information encoded in a qubit to a remote qudit? The rationale underlying this question are: (i) in order to transfer information encoded in a qubit to a remote qudit, the minimal requirement is  that the latter should have a nonzero vector polarisation in the spin operator basis that can be changed in a controlled manner, and, (ii) generation and manipulation of separable quantum states is easier than entangled quantum states. Interestingly, the answer to this question is in the affirmative. Such states have been identified in the study made in \cite{Bharath14}, in which the concept of equivalent states has been introduced.  An emergent feature of this concept is the notion of classical simulation of entangled states. To make the discussion easier, we have used the spin operator basis. The analysis is, however, applicable to any arbitrary system.

Two quantum states belonging to  Hilbert spaces of different dimensions are termed as {\it equivalent} if their Q-- representations are the same. Operationally, the crux of  the formalism involving {\it equivalent states} lies in the fact that the Q-- representation of a lower-dimensional noisy entangled state is the same as those of higher-dimensional noisy separable states. That is to say, the so-called lower-dimensional mixed entangled states admit classical simulation in higher dimensions. The mixed separable equivalent states of the two-qubit entangled Werner states have been obtained in \cite{Bharath14}. In a subsequent study \cite{Adhikary16}, the separable equivalents of $SU(2)$ invariant $3\times N$ states have been identified.  The higher dimensional equivalent states are not only separable but highly mixed also.
Thus, this formalism provides a niche for the processing of quantum information with highly mixed states as well.

In this paper, we propose various communication protocols, using higher-dimensional separable equivalents of $2 \times 2$ entangled Werner states, identified in \cite{Bharath14}. Motivated by the fact that information is encoded in a qubit in the form of  its Bloch vector, we propose protocols for remote transfer of information in a known and an unknown qubit to a remote qudit using $2\times N$ dimensional discordant states. We also propose a protocol for transferring information encoded in the polarization vector of a qudit to a remote qudit. Finally, we propose a protocol for swapping of quantum discord from two $2\times N$ dimensional discordant states to a $N\times N$ dimensional state. A ubiquitous feature of all these protocols is that none of them employ entangled states.

There has been considerable improvement in the generation and manipulation of higher dimensional separable orbital angular momentum states (see, e.g., \cite{Molina2007twisted, erhard2018twisted} and references therein). Thus, we believe that employing the protocols proposed in this paper, one can avoid the need of entangled states for certain quantum communication tasks.

The plan of the paper is as follows:   we briefly review the formalism to be employed in section (\ref{Formalism}). Thereafter, we turn our attention to applications of the formalism in section (\ref{Applications}), which is central to the paper. In sections (\ref{Remote state preparation unknown qubit}) and (\ref{Remote preparation known qubit}), the protocols for transfer of information encoded in an unknown and a known qubit to remote qudits are presented respectively. In section (\ref{Quantum Teleportation werner}), the protocol for transferring information from an unknown spin--$S$ state (having only nonzero vector polarisation) by employing $\frac{1}{2} \otimes S$ separable equivalent of a $\frac{1}{2}\otimes \frac{1}{2}$ entangled Werner state is  presented. In section (\ref{discord swapping}), the protocol for swapping of quantum discord from two $\frac{1}{2} \otimes S$ states to a single $S \otimes S$ state has been proposed.  Finally, we discuss future prospects and possible advantages of quantum communication with equivalent states in section (\ref{Generalisation}).
Section (\ref{Conclusion}) concludes the paper with closing remarks. 

\section{Formalism}
\label{Formalism}
In this section, we briefly recapitulate the formalism involving equivalent states. For a detailed discussion, one can refer to \cite{Bharath14}.

 \subsection{Spin coherent state and the Q-representation}
The spin-coherent state (SCS) $|\hat{n}(\theta, \phi)\rangle$ for a spin--$S$ particle may be generated by the action of the rotation group $(SU(2))$ on the state with the highest weight (stretched case), $|S_z=+S\rangle$ \cite{Radcliffe71},
 \begin{eqnarray}
 |\hat{n}(\theta, \phi)\rangle \equiv e^{-iS_z\phi} e^{-iS_y\theta} e^{-iS_z\psi}|S\rangle.
 \end{eqnarray}
Spin coherent states have the following properties:
 \begin{eqnarray}
 &\langle \hat{n}(\theta, \phi)|\hat{S} |\hat{n}(\theta, \phi)\rangle = \hat{n}(\theta, \phi),\nonumber\\
& |\langle \hat{n}|\hat{n}'\rangle|^2 = \Big(\dfrac{1+\hat{n}\cdot\hat{n}'}{2}\Big)^{2S},
\end{eqnarray}
where $\hat{S} = \frac{\vec{S}}{S}$. The set of all SCSs $\{|\hat{n}(\theta, \phi)\rangle\}$ forms an overcomplete set
\begin{eqnarray}
\dfrac{2S+1}{4\pi}\int \sin\theta d{\rm\theta} d{\rm\phi} |\hat{n}(\theta, \phi)\rangle\langle \hat{n}(\theta, \phi)| = \mathbb{1}.
\end{eqnarray}
Being overcomplete, SCSs allow any state to be {\it completely} expressed in terms of its diagonal elements
\begin{eqnarray}
F(\hat{n}) \equiv \dfrac{2S+1}{4\pi}\langle\hat{n}|\rho|\hat{n}\rangle.
\end{eqnarray} 
As an example, the Q-representation of a spin--$\dfrac{1}{2}$ state
\begin{eqnarray}
\rho = \dfrac{1}{2}(\mathbb{1}+\vec{\sigma}\cdot\vec{p}),
\end{eqnarray}
 is given by
\begin{eqnarray}
F(\hat{n})=  \dfrac{1}{4\pi}(1+\hat{n}\cdot\vec{p}).
\end{eqnarray}
Similarly, the Q-representation of a two-qubit state,
\begin{eqnarray}
\rho_{12}= \dfrac{1}{4}\big\{\mathbb{1}+\vec{\sigma}_1\cdot\vec{P}_1+\vec{\sigma}_2\cdot\vec{P}_2+\sigma_{1i}\sigma_{2j}\Pi_{ij}\big\},
\end{eqnarray}
is given by
\begin{eqnarray}
F(\hat{m}, \hat{n}) &&=\dfrac{4}{(4\pi)^2}\langle \hat{m}\otimes \hat{n}|\rho_{12}|\hat{m}\otimes \hat{n}\rangle\nonumber\\
&&\equiv \dfrac{1}{(4\pi)^2}\big\{1+\vec{P}_1\cdot\hat{m}+\vec{P}_2\cdot\hat{n}+\Pi_{ij}m_in_j\big\}.
\end{eqnarray}
In what follows, we show that   Q-representation serves to  identify equivalent states.
\subsection{Equivalent states}
\label{Equivalent states}
Since there is a bijective mapping between a state $\rho$ and its Q-- representation, the idea is to take a lower-dimensional state as an abstract entity and to look for all physical manifests (higher dimensional states) that yield the same expectation values for certain observables.

In order to elucidate the concept of equivalent states, let there be two Hilbert spaces, ${\cal H}^{d_1}$, $ {\cal H}^{d_2}$ of dimensions $d_1$ and $d_2$ respectively ($d_1 < d_2$). If a state $\rho \in {\cal H}^{d_1}$ has an equivalent state ${\rho}' \in {\cal H}^{d_2}$, it implies that for all the operators  $\hat{O} \in {\cal H}^{d_1}$, there exist operators $\hat{O}' \in {\cal H}^{d_2}$ such that
\begin{eqnarray}
{\rm Tr}({\rho}\hat{O}) = {\rm Tr}(\rho'\hat{O}').
\end{eqnarray}

 Operationally, equivalent states reproduce the same expectation values for all the observables, after suitable rescalings. 
 The equivalence does not extend to properties such as rank and purity of the state.  
 For example, the spin--$S$ equivalent of a spin--$\frac{1}{2}$ state, $\rho =\frac{1}{2}(\mathbb{1}+\vec{\sigma}\cdot\vec{P})$, is given by
 \begin{eqnarray}
 \label{Equivalent_qubit}
 \rho^S(\vec{P})=\dfrac{1}{2S+1}(\mathbb{1}+\hat{S}\cdot\vec{P}) \in {\cal H}^{2S+1};~ \hat{S} = \dfrac{\vec{S}}{S};~ S\neq 0.
 \end{eqnarray}
 The non-negativity of the eigenvalues mandates $|\vec{P}| \leq 1$, for all $S$. These states form a family of equivalent states all yielding the same probability distribution function, $F(\hat{n}) = \dfrac{1}{4\pi}(1+\vec{P}\cdot\hat{n})$. While for spin--$\frac{1}{2}$, $|\vec{P}|=1$ corresponds to a pure state, it is mixed for all other spins $S\neq \frac{1}{2}$. The crucial point is that all these states contain the same information which is encoded in their polarisation vector. We next discuss how to retrieve this information.
 
 \subsubsection{Retrieval of information from equivalent state of a qubit}
 \label{info}
 \noindent The observables required for retrieval of information encoded in $\rho^S(\vec{P})$ (given in equation (\ref{Equivalent_qubit})) are given by
 \begin{eqnarray}
 \label{Eqobservables}
     \hat{O}^S(\vec{P}) = \frac{3S}{S+1}\hat{S}\cdot\hat{P}.
 \end{eqnarray}
\noindent For $S=\frac{1}{2}$, the observable is simply given by $\vec{\sigma}\cdot\hat{P}$, whereas for higher $S$ values (close to 20), it is approximately given by $3\hat{S}\cdot\hat{P}$. 
An analogy may be drawn with the energy level splitting of a particle with a non-zero magnetic moment in the magnetic field (Zeeman effect). If the magnetic moment $\vec{\mu}$ is very small, a large enough magnetic field $\vec{B}$ will provide the same level splitting ($\propto \vec{\mu}\cdot\vec{B}$) as is obtained with a low magnetic field and higher magnetic moment.
  
\begin{center}
    \begin{figure}[!htb]
\centerline{\includegraphics[scale=0.2]{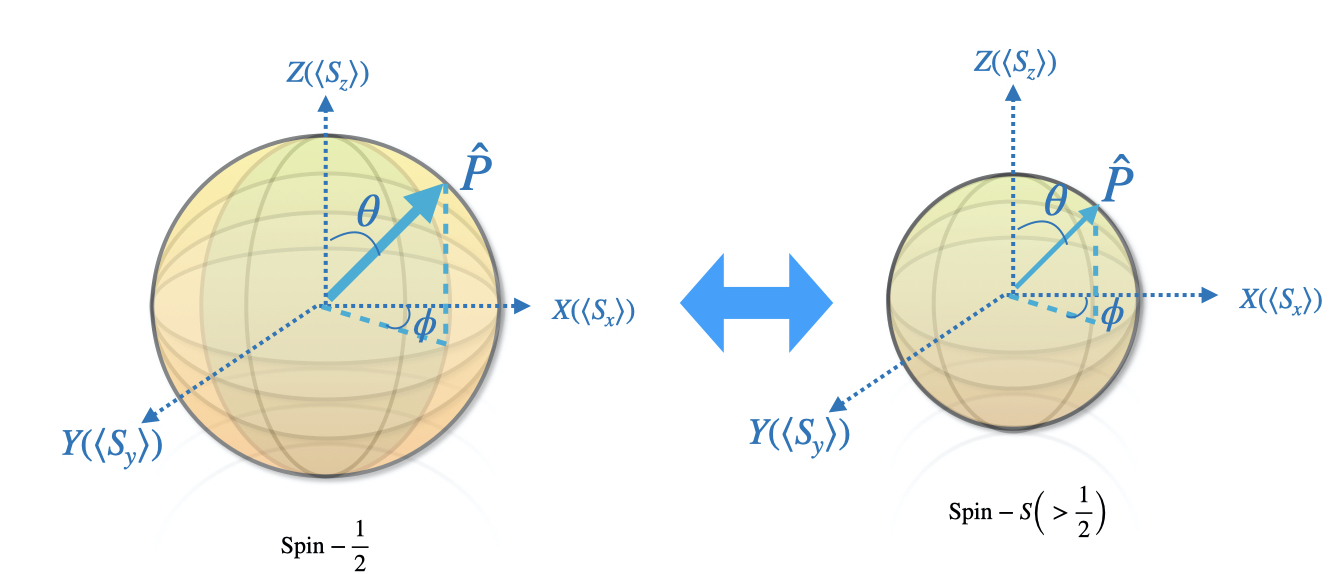}}
\caption{Pictorial representation of expectation values of $\hat{S}_x, \hat{S}_y, \hat{S}_z$ for a qubit and its spin--$S$ equivalent state. Note that the expectation values of $\hat{S}_x, \hat{S}_y, \hat{S}_z$ of spin--$S$ equivalent state fall short by a factor of $\frac{S+1}{3S}$, which is indicated by a reduced radius of the sphere. Thus, the equivalent observables get rescaled as given in equation (\ref{Eqobservables}).}
\label{Equivalence}
\end{figure}
\end{center}

  The equivalence of the expectation values of $\hat{S}_x, \hat{S}_y, \hat{S}_z$ of a spin--$\frac{1}{2}$ quantum state and of its spin--$S$ equivalent is shown pictorially in Fig. \ref{Equivalence}.
\subsection{Classical simulation of entangled states}
\label{Simulation}
With this concept of equivalent states, the emergent exquisite feature is the notion of classical simulation of entangled states \cite{Bharath14}. Classical simulation implies that a lower dimensional entangled state can be mimicked by  higher dimensional separable states, as they share the same Q-- representation. For example, consider the family of $2\times 2$ Werner states \cite{Werner89}
\begin{eqnarray}
\label{Werner}
\rho^{1/2\otimes 1/2}[\alpha]=\dfrac{1}{4}(\mathbb{1}-\alpha\vec{\sigma}_1\cdot\vec{\sigma}_2);~~ -\frac{1}{3}\leq\alpha\leq 1.
\end{eqnarray}
The $(2S_1+1)\times (2S_2+1)$-- dimensional equivalent state of $\rho^{1/2\otimes 1/2}[\alpha]$ is
\begin{eqnarray}
\label{Wernerequivalent}
\rho^{S_1\otimes S_2}[\alpha]= \dfrac{1}{(2S_1+1)(2S_2+1)}\big\{\mathbb{1}-\alpha\hat{S}_1\cdot\hat{S}_2\big\}.
\end{eqnarray}

The set of $2\times (2S+1)$ dimensional equivalent separable states of two-qubit entangled Werner states has been identified in \cite{Bharath14}. It has been shown that the state $\rho^{\frac{1}{2}\otimes S}[\alpha]$ is separable in the range $|\alpha| \leq \frac{S}{S+1}$. For any given $\alpha$, let $S_{\rm min}$ be the minimum value of $S$, for which the state $\rho^{\frac{1}{2}\otimes S}[\alpha]$ is separable. Its value is then given by the smallest half--integer greater than or equal to $\frac{|\alpha|}{1-|\alpha|}$.
Clearly, the value of $S_{\rm min}$ increases with an increase in the entanglement of two-qubit Werner states.

 Evidently, the separable equivalent of a pure two-qubit singlet state $(\alpha =1)$ does not exist in any finite dimension, so it is  termed as {\it exceptional}. We next show that the separable equivalent states of noisy entangled two-qubit Werner states are discordant.
 \subsection{Separable equivalent states of two--qubit Werner states have non--zero discord}
 \label{Nonzero_discord}
 The $2\times (2S+1)$-- dimensional equivalent state of the two-qubit Werner state $\rho^{\frac{1}{2}\otimes \frac{1}{2}}[\alpha] = \frac{1}{4}(\mathbb{1}-\alpha\vec{\sigma}_1\cdot\vec{\sigma}_2)$ is given by
 \begin{eqnarray}
 \label{Equivalent_Werner}
\rho^{\frac{1}{2}\otimes S}[\alpha] = \dfrac{1}{2(2S+1)}(\mathbb{1}-\alpha\vec{\sigma}_1\cdot\hat{S}_2).
 \end{eqnarray}
 The state $\rho^{\frac{1}{2}\otimes S}[\alpha]$ is separable in the region $|\alpha| \leq \frac{S}{S+1}$. We prove that it has a non-zero quantum discord in this range for $\alpha \neq 0$ by using its expansion in terms of separable states, given in \cite{Bharath14}.
 
 Consider the following state
 \begin{eqnarray}
     \rho_z \equiv \frac{1}{2}(\mathbb{1}-\beta\sigma_{1z})\otimes |S\rangle\langle S|;~ 0<\beta\leq 1.
 \end{eqnarray}
  In the irreducible tensor basis, it has the form
 \begin{eqnarray}
     \rho_z=\frac{1}{2}(\mathbb{1}-\beta\sigma_{1z})\otimes\dfrac{1}{2S+1}\Big\{\mathbb{1}+\frac{3S}{S+1}\hat{S}_{2z}+\sum_{k=2}^{2S}q_S^kS_{2z}^{(k)}\Big\}.
 \end{eqnarray}
Obviously,  the quantization axis has been chosen to be the $z$ axis.  The irreducible tensor operators are defined by, $S_{2z}^{(k)} = C_k(\vec{S}_2\cdot\vec{\nabla})^kr^kY_{k0}(\hat{r})$. The normalization factors, $C_k$, can be fixed conveniently. 
 
 Unlike $\rho^{\frac{1}{2}\otimes S}[\alpha]$, $\rho_z$ admits polarizations of all ranks $k \leq 2S$
and is anisotropic. We now construct similar states with the quantization axes along the $x$ and the $y$ directions and denote them by $\rho_{x,y}$, respectively. The separable state ${\bar \rho} =\frac{1}{3}(\rho_x + \rho_y + \rho_z)$ obtained by their incoherent superposition has the form
\begin{eqnarray}
    {\bar\rho} \equiv &&\dfrac{1}{3}(\rho_x +\rho_y +\rho_z)\nonumber\\
    =&& \dfrac{1}{2(2S+1)}\Big\{\mathbb{1}-\beta\dfrac{S}{S+1}\vec{\sigma}_1\cdot\hat{S}_2\Big\}+\cdots\nonumber\\
      =&& \dfrac{1}{2(2S+1)}\big\{\mathbb{1}-\alpha\vec{\sigma}_1\cdot\hat{S}_2\big\}+\cdots; ~~\alpha=\beta\dfrac{S}{S+1},
\end{eqnarray}
where the anisotropic terms are indicated by ellipses. To eliminate the unwanted anisotropic terms, we perform a uniformization by averaging over the full sphere, which of course leaves the isotropic terms unchanged, and yields
\begin{eqnarray}
    {\bar \rho}\rightarrow \dfrac{1}{4\pi}\int {\bar \rho}{\rm d\omega} \equiv \rho^{\frac{1}{2}\otimes S}[\alpha].
\end{eqnarray}
Evidently, the state $\rho^{\frac{1}{2}\otimes S}[\alpha]$ cannot be written as $\sum_i|\phi_{1i}\rangle\langle \phi_{1i}|\otimes \rho_{2i}$ or $\sum_j\rho_{1j}\otimes |\psi_{2j}\rangle\langle \psi_{2j}|$, where the sets $\{|\phi_{1i}\rangle\}$ and $\{|\psi_{2j}\rangle\}$ represent orthonormal bases in the spaces of the first and the second subsystems respectively. Thus,  $\rho^{\frac{1}{2}\otimes S}$ has a non--zero quantum discord, for all nonzero values of $\alpha$. 
\section{Applications}
\label{Applications}

In this section, we propose various applications which use $2\times N$-- dimensional separable \textit{equivalent} state $\rho^{1/2\otimes S}$, as a resource instead of the two-qubit entangled Werner state $\rho^{1/2\otimes 1/2}$. 
As stated before, information is encoded in a qubit in the form of its Bloch vector. So, to transfer the encoded information to a remote qudit, its vector polarisation (in the spin operator basis) has to be suitably changed. After this, the remote party may employ equivalent observables given in \cite{Bharath14} and also discussed in section (\ref{info}) to retrieve the encoded information. With this dictum, we propose various protocols as follows. The practical advantage that the proposed protocols serve is that of encoding information in quantum states quite close to the maximally mixed state\footnote{For $S=20$, the equivalent state $\frac{1}{2S+1}(\mathbb{1}+\frac{S_z}{S})$ (of a pure single-qubit state $\frac{1}{2}(\mathbb{1}+\sigma_z)$) has fidelity $0.876$ with the completely mixed state $\frac{1}{2S+1}\mathbb{1}$.}, which have hitherto been relegated from an information-theoretic viewpoint. We start with a protocol for remotely transferring information in an unknown qubit to a qudit.

\subsection{Transfer of information from an unknown qubit to a remote qudit}
\label{Remote state preparation unknown qubit}
 In this section, we present a protocol for the transfer of information in an unknown qubit to a remote qudit. 
 In the protocol, the separable state,
 \begin{eqnarray}
 \rho^{AB}(\alpha)^{1/2\otimes S} \equiv \dfrac{1}{2(2S+1)}(\mathbb{1}-\alpha\vec{\sigma}_2\cdot\hat{S}_3),
 \end{eqnarray}
 is shared between Alice and Bob and thus acts as a quantum channel for communication.  For a given $\alpha$, the minimum value of $S$, for which the state $\rho^{AB}(\alpha)^{1/2\otimes S}$ is separable, is given by the smallest half--integer greater than $\frac{|\alpha|}{1-|\alpha|}$.

Let $\rho_1^A=\frac{1}{2}(\mathbb{1}+\vec{\sigma}_1\cdot\vec{p})$ be the unknown qubit with Alice whose equivalent is to be remotely prepared.
 The superscripts of a state indicate the parties having or sharing the state, e.g., $\rho^{AB}$ represents a bipartite state shared between Alice and Bob. The protocol is as follows:
 \begin{enumerate}
     \item Alice has two qubits:
     \begin{enumerate}
         \item  a qubit, unknown to her, whose information is to be transferred remotely,
         \item the qubit of $\rho^{AB}(\alpha)^{1/2\otimes S}$, which has a shared correlation with the qudit at Bob.
     \end{enumerate}
  Thus, the combined state of Alice and Bob is given as \begin{equation}
      {\rho}^A_1\otimes \rho^{AB}(\alpha)^{1/2 \otimes S}=\frac{1}{2^2(2S+1)}(\mathbb{1}+\vec{\sigma}_1\cdot\vec{p})\otimes(\mathbb{1}-\alpha\vec{\sigma}_2\cdot\hat{S}_3).
  \end{equation}
  
\item  Alice performs a measurement in the Bell basis on her two qubits. She gets one of the four Bell states given below, with equal probability of $1/4$:
  \begin{eqnarray}
&  \rho^{AA}_{1} = \dfrac{1}{4}\big(\mathbb{1} -\vec{\sigma}_1\cdot\vec{\sigma}_2\big);~ &\rho^{AA}_{2} = \dfrac{1}{4}\big(\mathbb{1} -\sigma_{1x}\sigma_{2x}+\sigma_{1y}\sigma_{2y}+\sigma_{1z}\sigma_{2z}\big),\nonumber\\
  &  \rho^{AA}_{3}= \dfrac{1}{4}\big(\mathbb{1} +\sigma_{1x}\sigma_{2x}-\sigma_{1y}\sigma_{2y}+\sigma_{1z}\sigma_{2z}\big);~&  \rho^{AA}_{4}= \dfrac{1}{4}\big(\mathbb{1} +\sigma_{1x}\sigma_{2x}+\sigma_{1y}\sigma_{2y}-\sigma_{1z}\sigma_{2z}\big).\nonumber\\
  \end{eqnarray}
\item She sends the information about her measurement outcome through a classical channel to Bob. Depending upon the information received, Bob performs a corresponding rotation on his state to retrieve the equivalent state of the qubit.
\item If Alice obtains $\rho_1^{AA}, \rho_2^{AA}, \rho_3^{AA}, \rho_4^{AA}$, the transformations that Bob has to apply on his qudit are $\mathbb{1}, R_x(\pi), R_y(\pi), R_z(\pi)$ respectively. The symbols $R_x(\pi), R_y(\pi), R_z(\pi)$ represent Wigner rotation matrices of spin--$S$ about $x, y, $ and $z$ axes through an angle $\pi$. After performing the rotations,  the state at Bob is $\tilde{\rho}^B_1(\alpha) \equiv \frac{1}{2S+1}(\mathbb{1}+\alpha\hat{S_3}\cdot\hat{p})$. Different transformations to be applied by Bob depending on the measurement outcomes of Alice are summarised in the table (\ref{Table}).
 \end{enumerate}
 Thus, following the procedure, the spin--$S$ equivalent state of an unknown qubit is prepared at Bob's end. The calculations for all the four possibilities (in which Alice gets $\rho_1^{AA}, \cdots, \rho^{AA}_4$) are shown below.
 \begin{table}
 \begin{center}
\begin{tabular}{ |c|c|c| } 
 \hline
Post-measurement  & Collapsed state & Transformation to be\\ state of Alice &  of Bob &  applied by Bob \\\hline\hline 
 $\rho_1^{AA}$ & $\frac{1}{2S+1}(\mathbb{1}+\alpha\hat{S}_3\cdot\hat{p})$ & $\mathbb{1}$ \\
 \hline
 $\rho_2^{AA}$ & $\frac{1}{2S+1}(\mathbb{1}+\alpha(\hat{S}_{3x}p_{3x}-\hat{S}_{3y}p_{3y}-\hat{S}_{3z}p_{3z}))$ & $R_x(\pi)$ \\ 
 \hline
  $\rho_3^{AA}$ & $\frac{1}{2S+1}(\mathbb{1}+\alpha(-\hat{S}_{3x}p_{3x}+\hat{S}_{3y}p_{3y}-\hat{S}_{3z}p_{3z}))$ & $R_{y}(\pi)$ \\ 
  \hline
 $\rho_4^{AA}$ & $\frac{1}{2S+1}(\mathbb{1}+\alpha(-\hat{S}_{3x}p_{3x}-\hat{S}_{3y}p_{3y}+\hat{S}_{3z}p_{3z}))$ & $R_{z}(\pi)$  \\ 
 \hline
\end{tabular}
\caption{Post-measurement state of Alice and the respective transformation to be applied by Bob.}
\label{Table}
\end{center}
\end{table}
 \begin{center}
    \begin{figure}[!htb]
\centerline{\includegraphics[scale=0.2]{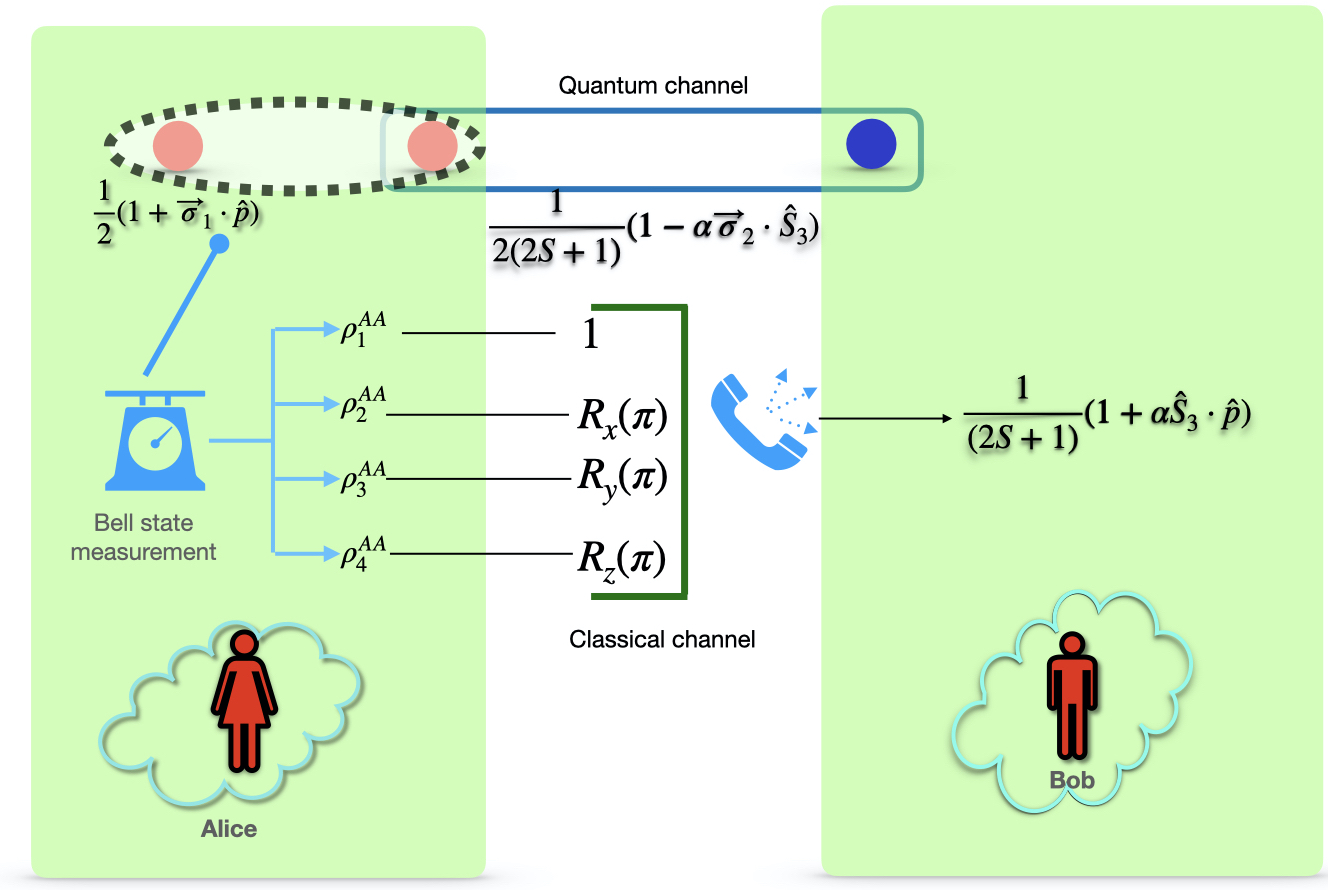}}
\caption{Pictorial representation of protocol for remote transfer of information from an unknown qubit to a qudit.}
\label{Remote_unknown_pic}
\end{figure}
\end{center}
 \subsubsection*{Calculation}
  We start with the case when the post-measurement state of Alice is the singlet state out of the four Bell states:
 \begin{eqnarray}
 \label{EquivalentBB}
&&\dfrac{1}{(2S+1)2^4} {\rm Tr}_{12}\Big\{\underbrace{\Big(\mathbb{1}-\vec{\sigma}_1\cdot\vec{\sigma}_2\Big)}_{\rm Measurement ~operator}\underbrace{(\mathbb{1}+\vec{\sigma}_1\cdot\hat{p})}_{\substack{{\rm Qubit~whose~polarisation}\\{\rm is~to~be~transferred~remotely}}}\otimes \underbrace{(\mathbb{1}-\alpha\vec{\sigma}_2\cdot\hat{S}_3)}_{\substack{{\rm State~shared~between}\\{\rm Alice~and~Bob}}}\Big\}\nonumber\\ =&&\dfrac{1}{4}\times\dfrac{1}{2S+1}(\mathbb{1}+\alpha\hat{S}_3\cdot\hat{p}).
\end{eqnarray}
Similarly, the post-measurement states corresponding to the
other three Bell states are shown below:
\begin{eqnarray}
 \label{EquivalentBB1}
\dfrac{1}{(2S+1)2^4}& {\rm Tr}_{12}\Big\{\Big(\mathbb{1}-(\sigma_{1x}\sigma_{2x}-\sigma_{1y}\sigma_{2y}-\sigma_{1z}\sigma_{2z})\Big)(\mathbb{1}+\vec{\sigma}_1\cdot\hat{p})\otimes (1-\alpha\vec{\sigma}_2\cdot\hat{S}_3)\Big\}\nonumber\\
&=\dfrac{1}{4}\times\dfrac{1}{2S+1} \big(\mathbb{1}+\alpha(\hat{S}_{3x}p_{3x} - \hat{S}_{3y}p_{3y} -\hat{S}_{3z}p_{3z})\big);
\end{eqnarray}
\begin{eqnarray}
 \label{EquivalentBB2}
\dfrac{1}{(2S+1)2^4}& {\rm Tr}_{12}\Big\{\Big(\mathbb{1}-(-\sigma_{1x}\sigma_{2x}+\sigma_{1y}\sigma_{2y}-\sigma_{1z}\sigma_{2z})\Big)(\mathbb{1}+\vec{\sigma}_1\cdot\hat{p})\otimes (1-\alpha\vec{\sigma}_2\cdot\hat{S}_3)\Big\}\nonumber\\
&=\dfrac{1}{4}\times\dfrac{1}{2S+1} \big(\mathbb{1}+\alpha(-\hat{S}_{3x}p_{3x} + \hat{S}_{3y}p_{3y} -\hat{S}_{3z}p_{3z})\big);
\end{eqnarray}
\begin{eqnarray}
\dfrac{1}{(2S+1)2^4}& {\rm Tr}_{12}\Big\{\Big(\mathbb{1}-(-\sigma_{1x}\sigma_{2x}-\sigma_{1y}\sigma_{2y}+\sigma_{1z}\sigma_{2z})\Big)(\mathbb{1}+\vec{\sigma}_1\cdot\hat{p})\otimes (1-\alpha\vec{\sigma}_2\cdot\hat{S}_3)\Big\}\nonumber\\
 \label{Equivalent_2}
&=\dfrac{1}{4}\times\dfrac{1}{2S+1} \big(\mathbb{1}+\alpha(-\hat{S}_{3x}p_{3x} - \hat{S}_{3y}p_{3y} +\hat{S}_{3z}p_{3z})\big).
 \end{eqnarray}
 Consequently, Bob will be in possession of the states given in the equations (\ref{EquivalentBB}),  (\ref{EquivalentBB1}),  (\ref{EquivalentBB2}) and  (\ref{Equivalent_2}) with equal probability of $\frac{1}{4}$.
The advantage of the proposed protocol is that it does not require an entangled state as  a quantum channel. Furthermore, mixed separable (but discordant) states act as a quantum channel which are relatively easier to produce.  Detailed calculations are given in Appendix (\ref{AppendixA}).


\subsection{Remote transfer of information of a known qubit to qudit}
\label{Remote preparation known qubit}
\begin{center}
    \begin{figure}[!htb]
\centerline{\includegraphics[scale=0.25]{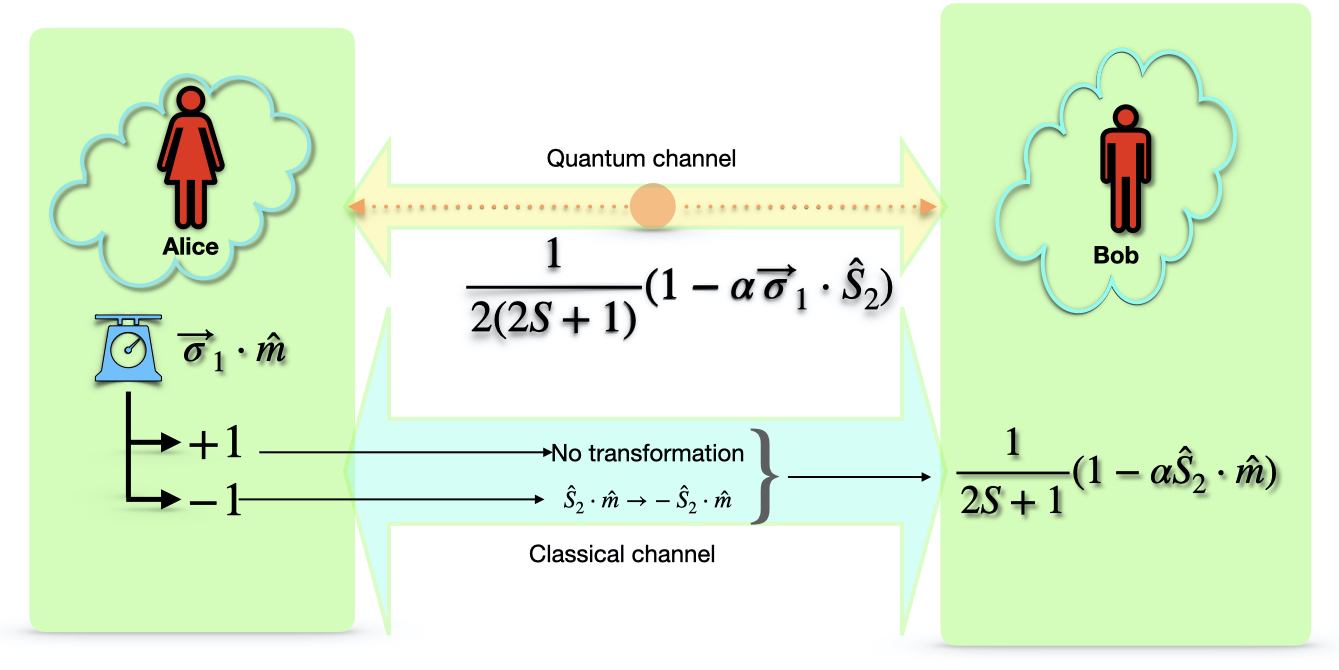}}
\caption{Pictorial representation of protocol for remote transfer of information from a known qubit to a qudit}
\label{Remote_known_pic}
\end{figure}
\end{center}

In this section, we lay down a protocol for remote transfer of information of a known qubit to a qudit. The quantum channel $\rho^{AB}(\alpha)$, as before, is the  $2 \times (2S+1)$ dimensional separable equivalent of $2 \times 2$-- entangled Werner state
\begin{eqnarray}
\rho^{AB}(\alpha) = \dfrac{1}{2(2S+1)}\big(\mathbb{1}-\alpha\vec{\sigma}_1\cdot\hat{S}_2\big).
\end{eqnarray}
Let us assume Alice wants to transfer the information  of her qubit $\frac{1}{2}(\mathbb{1}- \vec{\sigma_{1}}\cdot\hat{m})$ to Bob's qudit. The steps to be performed by Alice are as follows:

\begin{enumerate}
\item Alice measures $\vec{\sigma}_1\cdot\hat{m}$ on her qubit. She obtains either $+1$ or $-1$ with equal probability of $1/2$.
\item If she obtains $\pm 1$ as the measurement outcome, her subsystem collapses to the state $\pi_{\hat{m}}^{\pm}= \dfrac{1}{2}(\mathbb{1}\pm\vec{\sigma}_1\cdot\hat{m})$. The normalised post-measurement state of Bob's subsystem is
\begin{eqnarray}
\rho^B_{pm}= \frac{1}{2S+1}\big\{\mathbb{1}\mp \alpha\hat{S}_2\cdot\hat{m}\big\}.
\end{eqnarray}
\item If Alice obtains $+1$, she sends a classical message to Bob to do no transformation on his state. If Alice gets $-1$, she asks Bob to apply
a rotation of $\pi$ about an axis perpendicular to $\hat{m}$ (such that $\hat{S}_2\cdot\hat{m} \rightarrow -\hat{S}_2\cdot\hat{m}$). This is because the state $\frac{1}{(2S+1)}\big\{\mathbb{1}+\alpha\hat{S}_2\cdot\hat{m}\big\}$ can be changed to $\frac{1}{(2S+1)}\big\{\mathbb{1}-\alpha\hat{S}_2\cdot\hat{m}\big\}$ by this transformation. Thus, Alice can prepare the requisite state at Bob's end by sending one bit of classical information. 
\end{enumerate}
The pictorial representation of the proposed protocol is shown in Fig. \ref{Remote_known_pic}. 
\subsubsection{Evaluation of the performance of the protocols}
 Obviously, the performance of the protocols proposed in sections (\ref{Remote state preparation unknown qubit}) and (\ref{Remote preparation known qubit}) depends on the {\it closeness} of the qudit state prepared at Bob with the equivalent state of the  qubit at Alice. We employ two distance measures, {\it viz.}, fidelity, and Hilbert--Schmidt distance for analysing the performance  of the protocol.
\subsubsection*{Fidelity}
 Fidelity of two quantum states $\rho_1$ and $\rho_2$ is given by \cite{Nielsen00}
 \begin{eqnarray}
     F = \big({\rm Tr}\sqrt{\sqrt{\rho}_1\rho_2\sqrt{\rho}_1}\big)^2.
 \end{eqnarray}
 
\begin{center}
    \begin{figure}[!htb]
\centerline{\includegraphics[scale=0.3]{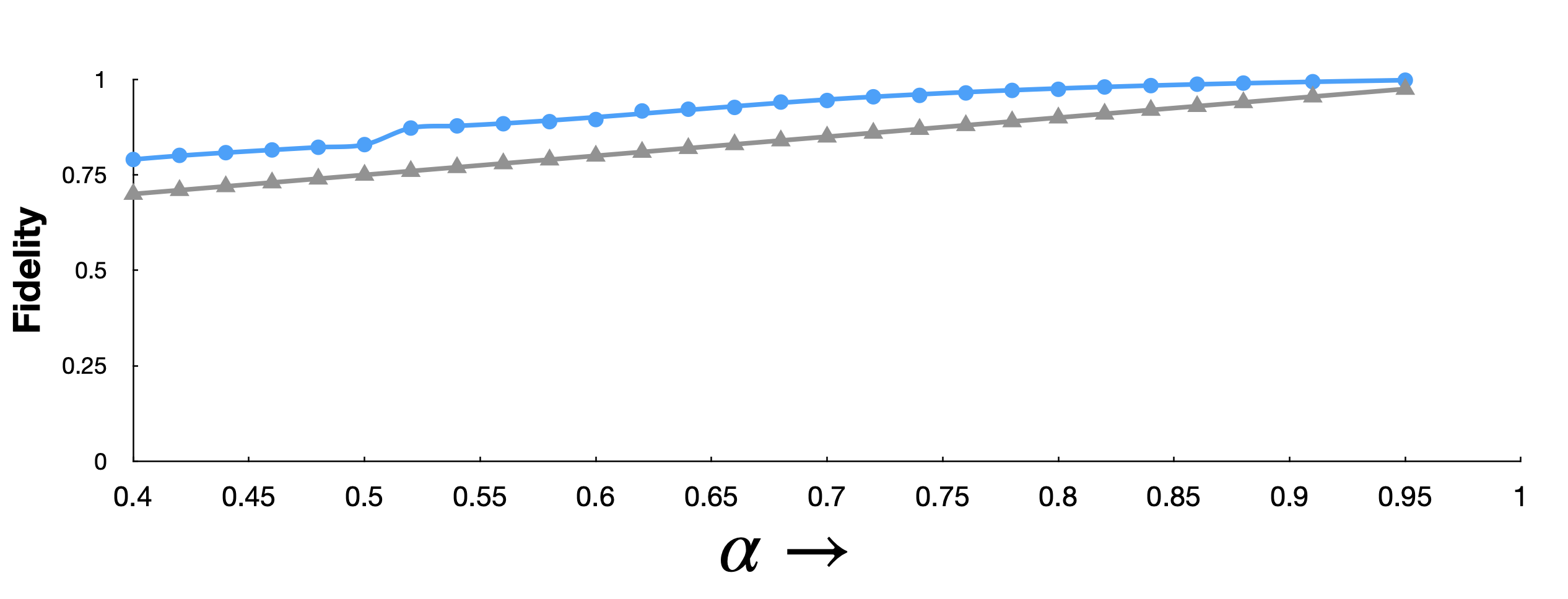}}
\caption{Blue: Fidelity of remotely prepared equivalent state w.r.t $\alpha$
for $S_{\rm{min}}$ 
when noisy separable channel is used. Grey: Fidelity of remotely prepared qubit with noisy channel for the same value of $\alpha$. Please note that for the case of qubits, the shared quantum channel is entangled.}
\label{fig1}
\end{figure}
\end{center}

In order to assess the performance of the protocols, we plot the fidelity of the remotely prepared qudit at Bob with the equivalent qudit of Alice's qubit as a function of $\alpha$ in  Fig. \ref{fig1}. For a given value of $\alpha$, we have taken the minimum value of $S$, corresponding to which the separable equivalent of the two-qubit Werner state exists (for details, see section (\ref{Simulation})). On the same graph, we also plot the fidelity of $\frac{1}{2}(\mathbb{1}+\sigma_z)$ with $\frac{1}{2}(\mathbb{1}+\alpha \sigma_z)$ that is prepared using an entangled two-qubit Werner state as a quantum channel in the quantum teleportation protocol \cite{Bennett93}.

Evidently, the plots show that it is easier to transfer information of a qubit to a remote qudit due to the following two reasons:
\begin{enumerate}
    \item For transfer of information from a qubit to a qudit, only a discordant state needs to be shared, and,
    \item For a given value of $\alpha$, the fidelity of the remotely prepared qudit with the equivalent qudit of the qubit (at Alice) is higher than that of the remotely prepared qubit with itself.
\end{enumerate}
\subsubsection*{Hilbert-Schmidt distance}
\begin{center}
    \begin{figure}[!htb]
\centerline{\includegraphics[scale=0.25]{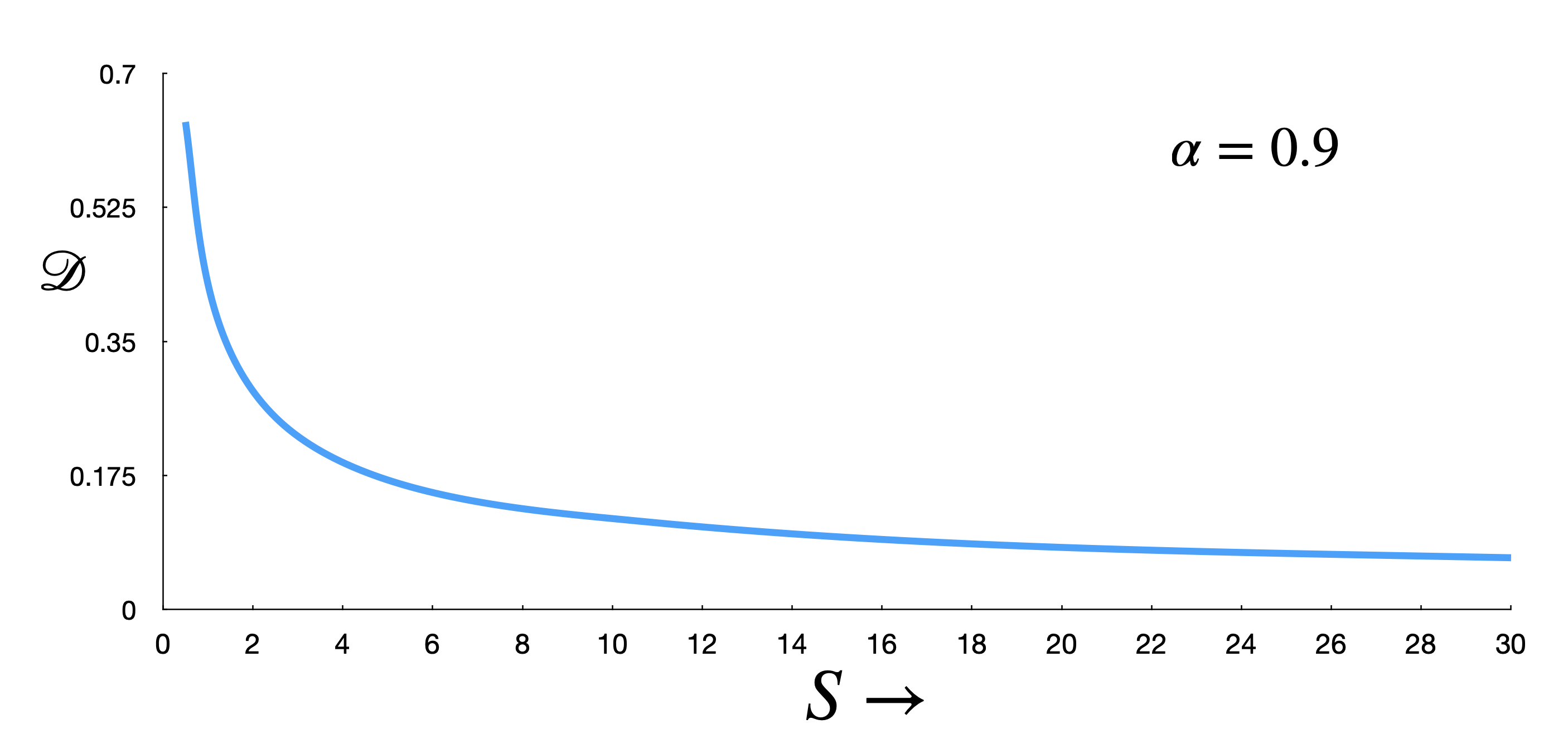}}
\caption{Plot of relative distance $\cal{D}$ with spin $S$ for $\alpha =0.9$. For $S \geq 9$, the channel is separable and for $S<9$, it is entangled (cf. section (\ref{Simulation})).}
\label{HilbertSchmidtPlot}
\end{figure}
\end{center}

The Hilbert--Schmidt distance between two quantum states $\rho_1$ and $\rho_2$ is given by \cite{Hillery96}
\begin{eqnarray}
    d=\sqrt{{\rm Tr}[(\rho_1-\rho_2)^2]}.
\end{eqnarray}
For the proposed protocol, we require the distance between $\rho_1=\frac{1}{2S+1}(\mathbb{1}+\hat{S}\cdot\hat{p})$ (equivalent qudit of Alice's qubit) and $\rho_2=\frac{1}{2S+1}(\mathbb{1}+\alpha\hat{S}\cdot\hat{p})$ (Bob's qudit at the end of the protocol), which is given as
\begin{eqnarray}
\label{HilbertSchmidt}
    d= (1-\alpha)\sqrt{\dfrac{S+1}{3S(2S+1)}}.
\end{eqnarray}
If the quantum channel between Alice and Bob is completely mixed,  Bob has a completely mixed state with him irrespective of the Bloch vector of Alice's qubit. In this case, the distance between (i) equivalent qudit of Alice's qubit, and, (ii) the qudit prepared at Bob, which is the maximally mixed state, is given as
\begin{eqnarray}
\label{HilbertSchmidtzero}
    d_0= \sqrt{\dfrac{S+1}{3S(2S+1)}}.
\end{eqnarray}
Thus, the difference of equations (\ref{HilbertSchmidtzero}) and (\ref{HilbertSchmidt}), termed as {\it relative distance},
\begin{eqnarray}
\label{HSdistance}
 {\cal D}\equiv  d_0-d=
  \alpha\sqrt{\dfrac{S+1}{3S(2S+1)}},
\end{eqnarray}
may be used for quantification of performance of the protocol. This is because it quantifies the reduction in the distance between the qudit prepared at Bob for a given value of $\alpha$, and, that prepared at Bob for $\alpha=0$ (which corresponds to the completely noisy channel). The derivative of the relative distance ${\cal D}$ (given in equation (\ref{HSdistance})) with respect to $S$ is
\begin{eqnarray}
\label{HSderivaitve}
    {\cal D}'(S) = -\frac{\alpha(2S^2+4S+1)}{2\sqrt{3(S+1)}S^{3/2}(2S+1)^{3/2}} <0,
\end{eqnarray}
which is negative for any nonzero $\alpha$. Clearly,  for a given $\alpha$, the relative distance ${\cal D}$ decreases as $S$ increases.  The plot of the relative distance ${\cal D}$ with $S$ is shown in Fig. \ref{HilbertSchmidtPlot} for $\alpha=0.9$. 

For large values of $S$, the asymptotic values of ${\cal D}$ and ${\cal D}'(S)$ (given in equations (\ref{HSdistance}) and (\ref{HSderivaitve}) respectively) are $\frac{\alpha}{\sqrt{6S}}$ and $-\frac{\alpha}{2\sqrt{6}S^{3/2}}$, both of which tend to zero as $S \to \infty$.

\subsubsection{Efficiency of the proposed protocols}
In this section, we show that the proposed protocols may be implemented efficiently, thanks to the separability of shared channels and availability of higher dimensional orbital angular momentum states. The yield, in these protocols, will be high  since there is no need for a nonlinear process to generate separable states.  In contrast, entangled states, which are generated by, say, spontaneous parametric down conversion, have a  very low yield with a reported efficiency of $4 \times 10^{-6}$ \cite{bock2016highly}. The techniques available for generation, manipulation and detection of orbital angular momentum states are well developed (see, for example, \cite{shen2019optical, padgett2017orbital} and references therein). Hence, they can be employed in the implementation of these protocols. Finally, apart from being merely discordant, the resource states employed in the protocols presented in this paper are mixed, which further reduces the burden of preparation.
\subsection{Transfer of information of an unknown qudit to a remote qudit with separable equivalent of Werner state}
\label{Quantum Teleportation werner}
In this section, we propose a protocol for the transfer of information of vector polarisation in the spin operator basis of an unknown qudit  to a remotely located qudit. As before, the $1/2 \otimes S$  separable equivalent of the $1/2 \otimes 1/2$ Werner state,
 \begin{eqnarray}
 \rho^{AB}(\alpha)^{1/2\otimes 1/2} \cong \rho^{AB}(\alpha)^{1/2\otimes S}=\frac{1}{2(2S+1)}(\mathbb{1}-\alpha\vec{\sigma}_2\cdot\hat{S}_3),
 \end{eqnarray}acts as a quantum channel in this protocol.
The steps of the protocol are as follows:
 \begin{enumerate}
     \item  Let $\rho^A=\frac{1}{2S+1}(\mathbb{1}+\hat{S}_1\cdot\vec{p})$ be the $(2S+1)$--dimensional state whose information is to be transferred by Alice. Thus, Alice has two subsystems; (i) the unknown state whose information she wants to transfer to a remote qudit, (ii) the other qubit which has shared correlation with Bob's qudit.  
     \item Alice performs a measurement of $\hat{S}_1\cdot\vec{\sigma}_2$, which is a dichotomic observable having eigenvalues $+1$ and $-\frac{S+1}{S}$. The two eigenvalues are degenerate with respective degeneracies of $2(S+1)$ and $2S$ respectively. The corresponding eigenprojections are
     \begin{eqnarray}
     && \Pi_1 =   \frac{S+1}{2S+1}\Big(\mathbb{1}+\frac{S}{S+1}\hat{S}_1\cdot\vec{\sigma}_2\Big),\nonumber\\
      &&\Pi_2= \frac{S}{2S+1}(\mathbb{1}-\hat{S_1}\cdot\vec{\sigma}_2).
       \end{eqnarray}
     \item Upon measurement of $\Pi_1$ and $\Pi_2$ by Alice, the respective post-measurement states at Bob are $\frac{1}{2S+1}\Big(\mathbb{1}-\frac{\alpha}{3}\hat{S}_3\cdot\vec{p}\Big)$ and $\frac{1}{2S+1}\Big(\mathbb{1}+\alpha\frac{S+1}{3S}\hat{S}_3\cdot\vec{p}\Big)$ with probabilities $\frac{S+1}{2S+1}$ and $\frac{S}{2S+1}$ respectively. 
     \item After performing the measurement, Alice sends Bob information about her measurement result. If Alice obtains $+1$, the direction of polarisation vector of Bob's qudit is opposite to that of Alice's qudit. On the other hand, if Alice gets $-\frac{S+1}{S}$, the direction of polarisation vector of Bob's qudit is the same as that of Alice's qudit. Since Bob has the knowledge of the channel, he knows that in both the cases, the polarisation vector of the qudit obtained by him has shrunk by a factor of approximately $\frac{1}{3}$ as compared to the qudit at Alice.  
\item As Bob is interested in the information encoded in the polarization vector, he needs to measure equivalent observables,  proposed in \cite{Bharath14}; given as $\frac{3S}{S+1}\hat{S}\cdot\hat{m}$, where $\hat{m}\in \{\hat{x}, \hat{y}, \hat{z}\}$. As explained in the protocol, the magnitude of the polarization vector is diminished by a constant factor $\frac{1}{3}$ which is a characteristic of the channel used. Therefore, it can be compensated by an extra factor of $3$ in the observable to be measured for retrieval of information (as explained in (\ref{info}) for the example of magnetic moment). Thus, Bob can retrieve full information about the polarization vector by measurement of appropriate observables.
 \end{enumerate}
 \begin{center}
    \begin{figure}[htb!]
\centerline{\includegraphics[scale=0.18]{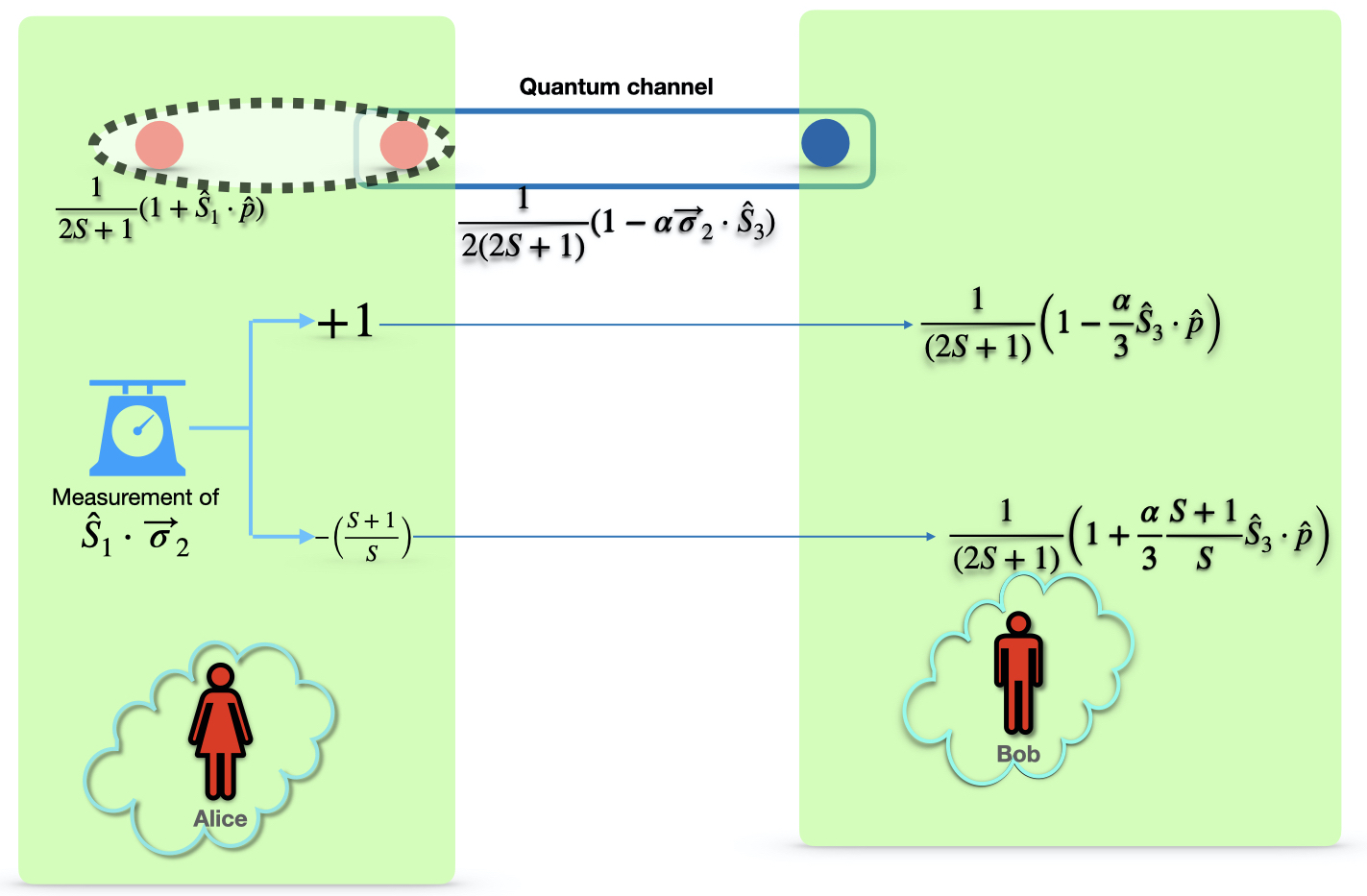}}
\caption{Pictorial representation of protocol for remote transfer of information of a qudit with $\frac{1}{2}\otimes S$ separable equivalent of a two-qubit entangled Werner state.}
\label{fig_eq}
\end{figure}
\end{center}
 The calculations for this are shown in appendix (\ref{Calculation_Teleportation}). The protocol is schematically shown in Fig. \ref{fig_eq}.

The advantage of this protocol is that in place of an entangled state, a mixed separable (but discordant) state acts as  a quantum channel. The above protocol can be used to send only those qudit states which have only non-zero vector polarization. However, this restriction can be relaxed if one uses the equivalent of higher dimensional entangled states.
\subsubsection{Evaluation of the performance of the protocol}
We employ the difference in the Hilbert Schmidt distances of remotely prepared qudits with a nonzero value of $\alpha$ and with $\alpha=0$ as the figure of merit. This difference, for the outcomes $+1$ and $-\frac{S+1}{S}$ for $\hat{S}_1\cdot\vec{\sigma}_2$, is given by the following expressions respectively:
\begin{eqnarray}
    {\cal D}_1 &&= \dfrac{\alpha}{3}\sqrt{\dfrac{S+1}{3S(2S+1)}},\nonumber\\
        {\cal D}_2 &&= \dfrac{\alpha (S+1)}{3S}\sqrt{\dfrac{S+1}{3S(2S+1)}}.
\end{eqnarray}
For large values of $S$, the asymptotic values of ${\cal D}_1$ and ${\cal D}_2$ are the same, given by $\frac{\alpha}{3\sqrt{6S}}$, both of which tend to $0$ as $S\to \infty$.
\subsection{Swapping of quantum discord}
 \label{discord swapping}
 Swapping of quantum correlations has been an interesting area of study, with the celebrated protocol proposed in \cite{Zukowski93} for entanglement swapping. Remote transfer of Gaussian quantum discord and swapping of quantum correlations between two-qubit Werner states have been earlier studied in \cite{Ma14} and \cite{Xie15} respectively.

 In this section, we describe a protocol for  swapping of quantum discord from $\frac{1}{2}\otimes S$ systems to $S \otimes S$ systems.  For this, let there be four parties named as Alice, Bob, Charlie and David. Let the states shared between the pairs (Alice, Bob) and (Charlie, David) be $\frac{1}{2(2S+1)}(\mathbb{1}-\alpha\vec{\sigma}_1\cdot\hat{S}_2)$ and $\frac{1}{2(2S+1)}(\mathbb{1}-\beta\hat{S}_3\cdot\vec{\sigma}_4)$ respectively. These states are separable equivalents of $\frac{1}{4}(\mathbb{1}-\alpha\vec{\sigma}_1\cdot\vec{\sigma}_2)$  and $\frac{1}{4}(\mathbb{1}-\beta\vec{\sigma}_3\cdot\vec{\sigma}_4)$ respectively. Note that the minimum value of $S$ gets fixed according to the values of $\alpha$ or $\beta$, as discussed in the section (\ref{Simulation}). Obviously, the states shared between the pairs (Alice, Bob) and (Charlie, David) are discordant states. There is no correlation between the pairs (Alice, David) and (Bob, Charlie). In this protocol, we show that quantum discord can be generated between Bob and Charlie, which are a-priori uncorrelated, by appropriate measurements.

\begin{center}
    \begin{figure}[!htb]
\centerline{\includegraphics[scale=0.18]{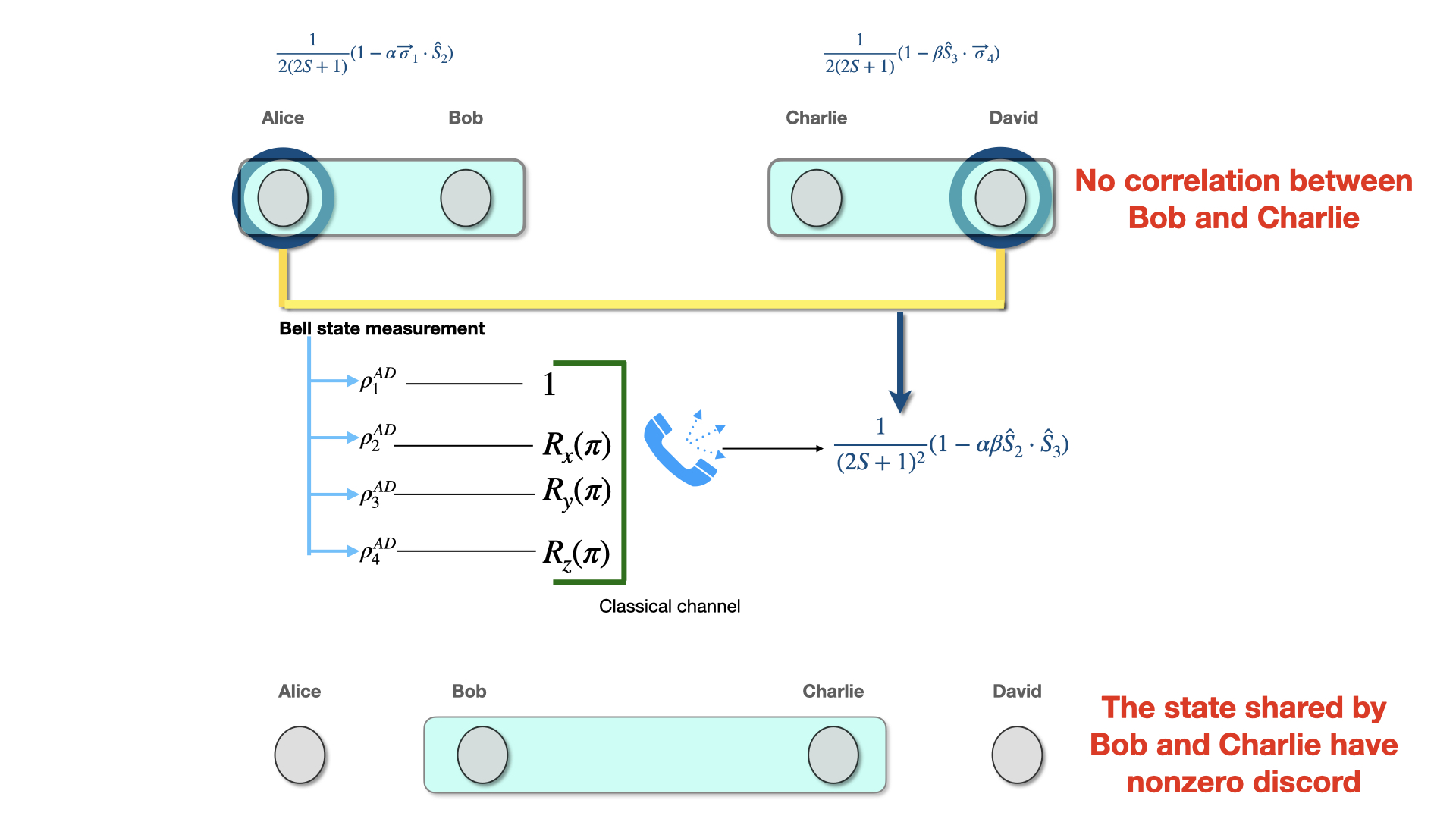}}
\caption{Pictorial representation of protocol for swapping of quantum discord.}
\label{Discord swapping}
\end{figure}
\end{center}

 The combined state of all the four-parties is as follows: 
\begin{eqnarray}
\dfrac{1}{\{2(2S+1)\}^2}(\mathbb{1}-\alpha\vec{\sigma}_1\cdot\hat{S}_2)(\mathbb{1}-\beta\hat{S}_3\cdot\vec{\sigma}_4).
\end{eqnarray}
If a measurement in the Bell basis is performed by Alice and David, the resultant state of Bob and Charlie will be discordant. For example, if a measurement of the singlet state is performed on the first and the fourth party, i.e., $\dfrac{1}{4}\{\mathbb{1}-\vec{\sigma}_1\cdot\vec{\sigma}_4\}$ is measured, then
\begin{eqnarray}
&&{\rm Tr}_{14}\dfrac{1}{2^4(2S+1)^2}\Big\{(\mathbb{1}-\vec{\sigma}_1\cdot\vec{\sigma}_4)(\mathbb{1}-\alpha\vec{\sigma}_1\cdot\hat{S}_2)(\mathbb{1}-\beta\hat{S}_3\cdot\vec{\sigma}_4)\Big\}\nonumber\\
=&&\dfrac{1}{2^4(2S+1)^2}{\rm Tr}_{14}\Big(\mathbb{1}-\alpha\beta\vec{\sigma}_1\cdot\hat{S}_2\vec{\sigma}_1\cdot\vec{\sigma}_4\hat{S}_3\cdot\vec{\sigma}_4\Big)\nonumber\\
=&&\dfrac{1}{4}\times\dfrac{1}{(2S+1)^2}(\mathbb{1}-\alpha\beta\hat{S}_2\cdot\hat{S}_3).
\end{eqnarray}
Similar calculations for the other three Bell states are shown in Appendix(\ref{Discord}).  The transformation to be applied by Bob corresponding to different measurements performed by Alice and David are shown in table (\ref{Table1}).
\begin{table}
 \begin{center}
\begin{tabular}{ |c|c|} 
 \hline
Measurement performed  & Transformation to be\\
by Alice and David &  applied by Bob \\\hline\hline 
$\rho_1^{AD}\equiv\frac{1}{4}(\mathbb{1}-\vec{\sigma}_1\cdot\vec{\sigma}_4)$ &  $\mathbb{1}$ \\
 \hline
$\rho_2^{AD}\equiv \frac{1}{4}(\mathbb{1}-{\sigma}_{1x}{\sigma}_{4x}+{\sigma}_{1y}{\sigma}_{4y}+{\sigma}_{1z}{\sigma}_{4z})$ & $R_x(\pi)$ \\ 
 \hline
$ \rho_3^{AD}\equiv  \frac{1}{4}(\mathbb{1}+{\sigma}_{1x}{\sigma}_{4x}+{\sigma}_{1y}{\sigma}_{4y}-{\sigma}_{1z}{\sigma}_{4z})$ &  $R_{z}(\pi)$ \\ 
  \hline
$ \rho_4^{AD}\equiv\frac{1}{4}(\mathbb{1}+{\sigma}_{1x}{\sigma}_{4x}-{\sigma}_{1y}{\sigma}_{4y}+{\sigma}_{1z}{\sigma}_{4z}))$ &  $R_{y}(\pi)$  \\ 
 \hline
\end{tabular}
\caption{Measurement performed by Alice and David and corresponding transformation to be applied by Bob.}
\label{Table1}
\end{center}
\end{table}
Thus, by having two lower-dimensional discordant states $\frac{1}{2(2S+1)}(\mathbb{1}-\alpha\vec{\sigma}_1\cdot\hat{S}_2)$ and  $\frac{1}{2(2S+1)}(\mathbb{1}-\beta\hat{S}_3\cdot\vec{\sigma}_4)$, a higher dimensional  discordant state $\frac{1}{(2S+1)^2}(\mathbb{1}-\alpha\beta\hat{S}_2\cdot\hat{S}_3)$ can be produced. The schematic representation of the protocol is given in Fig. \ref{Discord swapping}.
 
  The protocols proposed in sections (\ref{Remote state preparation unknown qubit}) and (\ref{discord swapping}) involve Bell state measurements. Fortunately, due to their vital role, Bell state measurements have been studied extensively \cite{Lutkenhaus99, Ghosh01, Calsamiglia02, Cao20, Fan21}. These studies would play a crucial role in the experimental implementation of the protocols.

 \subsection{Future prospects}
 \label{Generalisation}
 At this stage, it is worthwhile to consider the  future prospects of quantum communication with equivalent states and possible advantages that they would afford. Firstly, generalisations of our protocols to higher dimensional and multi-party systems constitute an interesting area of study. In fact, our protocols can be immediately generalised to remote transfer of information encoded in qutrits to remote qudits. This would require employment of the separable equivalents of $SU(2)$--invariant $3\times N$ level systems, as presented in \cite{Adhikary16}. Secondly, these protocols provide a motivation for the  identification of higher dimensional separable equivalent states of multi-party entangled systems, which is an interesting theoretical problem by itself. Those equivalent states can be employed for distribution of quantum information over a quantum network. In fact, a number of possibilities emerge when we seek to determine $k$--separable higher dimensional equivalent states  of an $N$--party lower dimensional fully entangled state.  The condition, $k=N$, implies a fully separable  state, whereas $k=1,\cdots, (N-1)$ represent partially entangled states, which are equivalent to a fully entangled state in lower dimensions.   Finally, this work shows the existence of quantum communication protocols in which we can dispense with  entanglement.
 
 \section{Conclusion}
 \label{Conclusion}
 In summary, this paper proposes various quantum communication protocols using $\frac{1}{2}\otimes S$ discordant states, identified in \cite{Bharath14}, as separable equivalents of entangled $\frac{1}{2}\otimes \frac{1}{2}$ Werner states. We have shown that remote transfer of equivalent state of qubit can be performed with approximately the same fidelity as noisy entangled state. Thus, the requirement of noisy entangled states, whose generation is very difficult, can be removed for transfer of information of polarization vector in the spin operator basis. We have also shown how  quantum discord in higher-dimensional states can be generated from two lower-dimensional discordant states by appropriate measurements.
 
 Finally, the protocols based on separable equivalents of two-qutrit entangled states constitute an interesting study which will be undertaken elsewhere.

 \acknowledgement
 It is a pleasure to thank Soumik Adhikary for discussions. We thank the anonymous referees, whose comments have really helped us improve the quality of the presentation and for  bringing several relevant references to our attention. Sooryansh and Rajni thank  CSIR (Grant no.: 09/086 (1278)/ 2017-EMR-I) and UGC for funding their research respectively.
 \section*{Author contribution statement}
 All the authors  have contributed equally in all respects.
 \begin{appendix}
 \section{Transfer of information from an unknown qubit to a remote qudit}
 \label{AppendixA}
 Detailed calculations leading to equation (\ref{EquivalentBB}), (\ref{EquivalentBB1}), (\ref{EquivalentBB2}) and (\ref{Equivalent_2}) are as follows. They essentially employ three properties, {\it viz.}, tracelessness of Pauli matrices, ${\rm Tr}(\vec{\sigma}\cdot\hat{m}\vec{\sigma}\cdot\hat{n}) = 2\hat{m}\cdot\hat{n}$, and,  ${\rm Tr}(A\otimes B)={\rm Tr}(A){\rm Tr}(B).$ It is sufficient to illustrate it for one case. The same method works for the other three equations as well.

 \noindent We start with the case when Alice measures a singlet state out of the four Bell states, i.e., $\rho_1^{AA}=\frac{1}{4}(\mathbb{1}-\vec{\sigma}_1\cdot\vec{\sigma}_2)$, which leads to  equation (\ref{EquivalentBB}), as follows:
 \begin{eqnarray}
&&\dfrac{1}{(2S+1)2^4} {\rm Tr}_{12}\Big\{\Big(\mathbb{1}-\vec{\sigma}_1\cdot\vec{\sigma}_2\Big)(\mathbb{1}+\vec{\sigma}_1\cdot\hat{p})\otimes(1-\alpha\vec{\sigma}_2\cdot\hat{S}_3)\Big\}\nonumber\\
&&=\dfrac{1}{(2S+1)2^4} {\rm Tr}_{12}\Big\{\Big(\mathbb{1}-\vec{\sigma}_1\cdot\vec{\sigma}_2\Big)\Big(\mathbb{1}+\vec{\sigma}_1\cdot\hat{p}-\alpha \vec{\sigma}_2\cdot\hat{S}_3-\alpha \vec{\sigma_1}\cdot\hat{p}\otimes\vec{\sigma_2}\cdot\hat{S}_3\Big)\Big\}\nonumber\\
&&=\dfrac{1}{(2S+1)2^4} {\rm Tr}_{12}\Big\{\mathbb{1}+\vec{\sigma_1}\cdot\hat{p}-\alpha \vec{\sigma}_2\cdot\hat{S}_3-\alpha \vec{\sigma}_1\cdot\hat{p}\otimes\vec{\sigma}_2\cdot\hat{S}_3\nonumber\\
&&~-\sigma_{1x}\sigma_{2x}(\mathbb{1}+\vec{\sigma}_1\cdot\hat{p}-\alpha \vec{\sigma}_2\cdot\hat{S}_3-\alpha \vec{\sigma}_1\cdot\hat{p}\otimes\vec{\sigma}_2\cdot\hat{S}_3)-\sigma_{1y}\sigma_{2y}(\mathbb{1}+\vec{\sigma}_1\cdot\hat{p}\nonumber\\
&&-\alpha \mathbb{1}\otimes\vec{\sigma}_2\cdot\hat{S}_3-\alpha \vec{\sigma}_1\cdot\hat{p}\otimes\vec{\sigma}_2\cdot\hat{S}_3)-\sigma_{1z}\sigma_{2z}(\mathbb{1}+\vec{\sigma}_1\cdot\hat{p}-\alpha \vec{\sigma}_2\cdot\hat{S}_3-\alpha \vec{\sigma}_1\cdot\hat{p}\otimes\vec{\sigma}_2\cdot\hat{S}_3)\Big\}\nonumber\\
&&=\dfrac{1}{(2S+1)2^4}\Big\{4\mathbb{1}+4p_x\hat{S}_{3x}+4p_y\hat{S}_{3y}+4p_z\hat{S}_{3z}\Big\}\nonumber\\
&&=\dfrac{1}{4}\times\dfrac{1}{2S+1}(\mathbb{1}+\alpha\hat{S}_3\cdot\hat{p}),
\end{eqnarray}
which is the same as the equation (\ref{EquivalentBB}).
\section{Transfer of information from an unknown qudit to a remote qudit}
\label{Calculation_Teleportation}
 When Alice measures $\hat{S}_1\cdot\vec{\sigma}_2$, she gets the states $\dfrac{1}{2S+1}\Big(\mathbb{1}-\dfrac{\alpha}{3}\hat{S}_3\cdot\vec{p}\Big)$ and $\dfrac{1}{2S+1}\Big(\mathbb{1}+\dfrac{\alpha}{3}\dfrac{S+1}{S}\hat{S}_3\cdot\vec{p}\Big)$ with respective probabilities of $\frac{S+1}{2S+1}$ and $\frac{S}{2S+1}$. The detailed calculations are as follows, which essentially employ the following two properties:
 \begin{enumerate}
     \item The spin operators $S_x, S_y, S_z$ are traceless, i.e., ${\rm Tr}~ S_i=0$ and ${\rm Tr}~ \sigma_j=0$, $i, j \in {x, y, z}$.
     \item  Trace of tensor product of two operators is equal to the product of the trace of operators, i.e., ${\rm Tr}(A\otimes B)={\rm Tr}(A){\rm Tr}(B).$
 \end{enumerate}
 \textbf{Case I:} When Alice's state is projected into the eigenstate corresponding to the eigenvalue +1, which occurs with a probability $\frac{S+1}{2S+1}$, then
 \begin{eqnarray}
&&\dfrac{1}{2(2S+1)^2} {\rm Tr}_{12}\Big\{{\Big(\mathbb{1}+\dfrac{S}{S+1}\hat{S}_1\cdot\vec{\sigma}_{2}\Big)}(\mathbb{1}+\hat{S}_1\cdot\vec{p})\otimes (\mathbb{1}-\alpha\vec{\sigma}_2\cdot\hat{S}_3)\Big\}\nonumber\\
&&=\dfrac{1}{2(2S+1)^2}{\rm Tr}_{12}\Big\{{\Big(\mathbb{1}+\dfrac{S}{S+1}\hat{S}_1\cdot\vec{\sigma}_{2}\Big)}\Big(\mathbb{1}+\hat{S}_1\cdot\vec{p}-\alpha\vec{\sigma}_2\cdot\hat{S}_3-\alpha\hat{S}_1\cdot\vec{p}\otimes \vec{\sigma}_2\cdot\hat{S}_3 \Big)\Big\}\nonumber\\
&&=\dfrac{1}{2(2S+1)^2}{\rm Tr}_{12}\Big\{\mathbb{1}+\hat{S}_1\cdot\vec{p}-\alpha\vec{\sigma}_2\cdot\hat{S}_3-\alpha\hat{S}_1\cdot\vec{p}\otimes \vec{\sigma}_2\cdot\hat{S}_3+\dfrac{S}{S+1}\hat{S}_1\cdot\vec{\sigma}_{2}\nonumber\\
&&+\dfrac{S}{S+1}(\hat{S}_1\cdot\vec{\sigma}_{2})(\hat{S}_1\cdot\vec{p})-\alpha\dfrac{S}{S+1}(\hat{S}_1\cdot\vec{\sigma}_{2})(\vec{\sigma}_2\cdot\hat{S}_3)-\alpha\dfrac{S}{S+1}(\hat{S}_1\cdot\vec{\sigma}_{2})(\hat{S}_1\cdot\vec{p}\otimes \vec{\sigma}_2\cdot\hat{S}_3) \Big\}\nonumber\\
&&=\dfrac{1}{2(2S+1)^2}\Big\{\mathbb{1}+\hat{S}_1\cdot\vec{p}-\alpha{ \vec{\sigma}_2\cdot\hat{S}_3}-\alpha(\hat{S}_1\cdot\vec{p})\otimes \vec{\sigma}_2\cdot\hat{S}_3+\dfrac{S}{S+1}\hat{S}_1\cdot\vec{\sigma}_{2}\nonumber\\
&&+\dfrac{S}{S+1}(\hat{S}_1\cdot\vec{\sigma}_{2})(\hat{S}_1\cdot\vec{p}-\alpha\dfrac{S}{S+1}(\hat{S}_1\cdot\vec{\sigma}_{2})(\vec{\sigma}_2\cdot\hat{S}_3)-\alpha\dfrac{S}{S+1}(\hat{S}_1\cdot\vec{\sigma}_{2})(\hat{S}_1\cdot\vec{p}\otimes \vec{\sigma}_2\cdot\hat{S}_3) \Big\}\nonumber\\
&&=\dfrac{1}{2S+1}\Big(\mathbb{1}-\dfrac{\alpha}{3}\hat{S}_3\cdot\vec{p}\Big).
\end{eqnarray}
The formula that we have used ${\rm Tr}(\hat{S}_i\hat{S}_j) = \frac{(S+1)(2S+1)}{3S}\delta_{ij}$ can be proved as follows. We have
\begin{eqnarray}
    S_x^2+S_y^2+S_z^2 = S(S+1)\mathbb{1}.
\end{eqnarray}
Taking trace of both sides and employing the property that ${\rm Tr}S_x^2 = {\rm Tr}S_y^2 = {\rm Tr}S_z^2$, we obtain
\begin{eqnarray}
    3{\rm Tr}S_x^2 &&= S(S+1){\rm Tr}\mathbb{1} = S(S+1)(2S+1)\nonumber\\
   \implies {\rm Tr}S_x^2 &&= \frac{S(S+1)(2S+1)}{3}.
\end{eqnarray}
Similarly, ${\rm Tr}S_y^2 = {\rm Tr}S_z^2 =  \frac{S(S+1)(2S+1)}{3}$. Furthermore, we have the following expression of invariance of trace under a unitary transformation,
\begin{eqnarray}
    {\rm Tr}(S_xS_y) = {\rm Tr}(US_xS_yU^{\dagger}),
\end{eqnarray}
where $U$ is a unitary transformation. Consider a rotation about the $X$ axis such that $\hat{y} \rightarrow -\hat{y}$ and $\hat{z}\rightarrow -\hat{z}$. Thus, under the effect of this transformation
\begin{eqnarray}
    {\rm Tr}(S_xS_y) = {\rm Tr}(-S_xS_y)\implies {\rm Tr}(S_xS_y) =0.
    \end{eqnarray}
    Both of these results can be combined to the following equation
\begin{eqnarray}\label{eq:trace}
    {\rm Tr}(\hat{S}_i\hat{S}_j) = \frac{(S+1)(2S+1)}{3S}\delta_{ij}.
\end{eqnarray}    
\textbf{Case II:}  When Alice's state is projected to the eigenstate corresponding to the eigenvalue $-\frac{S+1}{S}$, which occurs with a probability $\frac{S}{2S+1}$, then
 \begin{eqnarray}
&\dfrac{1}{2(2S+1)^2}{\rm Tr}_{12}\Big\{(\mathbb{1}-\hat{S}_1\cdot\vec{\sigma}_{2})(\mathbb{1}+\hat{S}_1\cdot\vec{p})\otimes (1-\alpha\vec{\sigma}_2\cdot\hat{S}_3)\Big\} =\dfrac{1}{2S+1}\Big(\mathbb{1}+\alpha\dfrac{(S+1)}{3S}\hat{S}_3\cdot\vec{p}\Big).\nonumber\\
\end{eqnarray}
This result is calculated following the same methodology as done for  the Case I.
 \section{Swapping of quantum discord}
 \label{Discord}
In this appendix, we present those cases of swapping of quantum discord, proposed in section (\ref{discord swapping}), in which the remaining three Bell states are measured. The calculations for rest of the three projection operators of Bell basis are as follows. They essentially employ three properties, {\it viz.}, tracelessness of Pauli matrices, ${\rm Tr}(\vec{\sigma}\cdot\hat{m}\vec{\sigma}\cdot\hat{n}) = 2\hat{m}\cdot\hat{n}$, and,  ${\rm Tr}(A\otimes B)={\rm Tr}(A){\rm Tr}(B).$  
 \begin{enumerate}
     \item When the measurement of $\dfrac{1}{4}\Big(\mathbb{1}-({\sigma}_{1x}{\sigma}_{4x}-{\sigma}_{1y}{\sigma}_{4y}-{\sigma}_{1z}{\sigma}_{4z})\Big)$ is made, the post-measurement state is
 \begin{eqnarray}
&&{\rm Tr}_{14}\dfrac{1}{2^4(2S+1)^2}\Big\{\Big(\mathbb{1}-({\sigma}_{1x}{\sigma}_{4x}-{\sigma}_{1y}{\sigma}_{4y}-{\sigma}_{1z}{\sigma}_{4z})\Big)(\mathbb{1}-\alpha\vec{\sigma}_1\cdot\hat{S}_2)(\mathbb{1}-\beta\hat{S}_3\cdot\vec{\sigma}_4)\Big\}\nonumber\\
&&=\dfrac{1}{2^4(2S+1)^2}{\rm Tr}_{14}\Big(\mathbb{1}-\alpha\beta\vec{\sigma}_{1}\cdot\hat{S}_2({\sigma}_{1x}{\sigma}_{4x}-{\sigma}_{1y}{\sigma}_{4y}-{\sigma}_{1z}{\sigma}_{4z})\hat{S}_3\cdot\vec{\sigma}_4\Big)\nonumber\\
&&=\dfrac{1}{2^4(2S+1)^2}{\rm Tr}_{14}\Big(\mathbb{1}-\alpha\beta\vec{\sigma}_{1}\cdot\hat{S}_2({\sigma}_{1x}{\sigma}_{4x}\hat{S_{3x}}\sigma_{4x}+\sigma_{1x}{\sigma}_{4x}\hat{S}_{3y}\sigma_{4y}+\sigma_{1x}{\sigma}_{4x}\hat{S}_{3z}\sigma_{4z}\nonumber\\
&&-{\sigma}_{1y}{\sigma}_{4y}\hat{S}_{3x}\sigma_{4x}-{\sigma}_{1y}{\sigma}_{4y}\hat{S}_{3y}\sigma_{4y}-{\sigma}_{1y}{\sigma}_{4y}\hat{S}_{3z}\sigma_{4z}-{\sigma}_{1z}{\sigma}_{4z}\hat{S}_{3x}\sigma_{4x}\nonumber\\
&&-{\sigma}_{1z}{\sigma}_{4z}\hat{S}_{3y}\sigma_{4y}-{\sigma}_{1z}{\sigma}_{4z}\hat{S}_{3z}\sigma_{4z}\Big)\nonumber\\
&&=\dfrac{1}{2^4(2S+1)^2}{\rm Tr}_{1}\Big(2\mathbb{1}-\alpha\beta\vec{\sigma}_{1}\cdot\hat{S}_2({2\sigma}_{1x}\hat{S}_{3x}-2{\sigma}_{1y}\hat{S}_{3y}-2{\sigma}_{1z}\hat{S}_{3z}\Big)\nonumber\\
&&=\dfrac{1}{2^3(2S+1)^2}{\rm Tr}_{1}\Big(\mathbb{1}-\alpha\beta\vec{\sigma}_{1}\cdot\hat{S}_2({\sigma}_{1x}\hat{S}_{3x}-{\sigma}_{1y}\hat{S}_{3y}-{\sigma}_{1z}\hat{S}_{3z}\Big)\nonumber\\
&&=\dfrac{1}{2^3(2S+1)^2}{\rm Tr}_{1}\Big(\mathbb{1}-\alpha\beta(\sigma_{1x}\hat{S}_{2x}({\sigma}_{1x}\hat{S}_{3x}-{\sigma}_{1y}\hat{S}_{3y}-{\sigma}_{1z}\hat{S}_{3z})+\sigma_{1y}\hat{S}_{2y}({\sigma}_{1x}\hat{S}_{3x}-{\sigma}_{1y}\hat{S}_{3y}-{\sigma}_{1z}\hat{S}_{3z})\nonumber\\
&&+\sigma_{1z}\hat{S}_{2z}({\sigma}_{1x}\hat{S}_{3x}-{\sigma}_{1y}\hat{S}_{3y}-{\sigma}_{1z}\hat{S}_{3z})\Big)\nonumber\\
&&=\dfrac{1}{2^3(2S+1)^2}{\rm Tr}_{1}\Big(\mathbb{1}-\alpha\beta(\sigma_{1x}\hat{S}_{2x}({\sigma}_{1x}\hat{S}_{3x}-{\sigma}_{1y}\hat{S}_{3y}-{\sigma}_{1z}\hat{S}_{3z})\nonumber\\
&&~~~+\sigma_{1y}\hat{S}_{2y}({\sigma}_{1x}\hat{S}_{3x}-{\sigma}_{1y}\hat{S}_{3y}-{\sigma}_{1z}\hat{S}_{3z})+\sigma_{1z}\hat{S}_{2z}({\sigma}_{1x}\hat{S}_{3x}-{\sigma}_{1y}\hat{S}_{3y}-{\sigma}_{1z}\hat{S}_{3z})\Big)\nonumber\\
&&=\dfrac{1}{2^3(2S+1)^2}\Big(2\mathbb{1}-\alpha\beta(2\hat{S}_{2x}\hat{S}_{3x}-2\hat{S}_{2y}\hat{S}_{3y}-2\hat{S}_{3z}\hat{S}_{3z})\Big)\nonumber\\
&&=\dfrac{1}{4}\times\dfrac{1}{(2S+1)^2}\Big(\mathbb{1}-\frac{\alpha\beta}{S^2}(S_{2x}S_{3x}-S_{2y}S_{3y}-S_{2z}S_{3z})\Big).
\end{eqnarray}
The factor of $\frac{1}{4}$ represents the probability with which this state is prepared. Other calculations are performed using the same approach.
  \item When the measurement of $\dfrac{1}{4}\Big(\mathbb{1}-(-{\sigma}_{1x}{\sigma}_{4x}-{\sigma}_{1y}{\sigma}_{4y}+{\sigma}_{1z}{\sigma}_{4z})\Big)$ is made
\begin{eqnarray}
&&{\rm Tr}_{14}\dfrac{1}{2^4(2S+1)^2}\Big\{\Big(\mathbb{1}-(-{\sigma}_{1x}{\sigma}_{4x}-{\sigma}_{1y}{\sigma}_{4y}+{\sigma}_{1z}{\sigma}_{4z})\Big)(\mathbb{1}-\alpha\vec{\sigma}_1\cdot\hat{S}_2)(\mathbb{1}-\beta\hat{S}_3\cdot\vec{\sigma}_4)\Big\}\nonumber\\
&&=\dfrac{1}{2^4(2S+1)^2}{\rm Tr}_{14}\Big(\mathbb{1}-\alpha\beta\vec{\sigma}_{1}\cdot\hat{S}_2(-{\sigma}_{1x}{\sigma}_{4x}-{\sigma}_{1y}{\sigma}_{4y}+{\sigma}_{1z}{\sigma}_{4z})\hat{S}_3\cdot\vec{\sigma}_4\Big)\nonumber\\
&&=\dfrac{1}{4}\times\dfrac{1}{(2S+1)^2}\Big(\mathbb{1}-\frac{\alpha\beta}{S^2}(-S_{2x}S_{3x}-S_{2y}S_{3y}+S_{2z}S_{3z})\Big).
\end{eqnarray}
The factor of $\frac{1}{4}$ represents the probability with which this state is prepared.
  \item When the measurement of $\dfrac{1}{4}\Big(\mathbb{1}-(-{\sigma}_{1x}{\sigma}_{4x}+{\sigma}_{1y}{\sigma}_{4y}-{\sigma}_{1z}{\sigma}_{4z})\Big)$ is made
\begin{eqnarray}
&&{\rm Tr}_{14}\dfrac{1}{2^4(2S+1)^2}\Big\{\Big(\mathbb{1}-(-{\sigma}_{1x}{\sigma}_{4x}+{\sigma}_{1y}{\sigma}_{4y}-{\sigma}_{1z}{\sigma}_{4z})\Big)(\mathbb{1}-\alpha\vec{\sigma}_1\cdot\hat{S}_2)(\mathbb{1}-\beta\hat{S}_3\cdot\vec{\sigma}_4)\Big\}\nonumber\\
&&=\dfrac{1}{2^4(2S+1)^2}{\rm Tr}_{14}\Big(\mathbb{1}-\alpha\beta\vec{\sigma}_{1}\cdot\hat{S}_2(-{\sigma}_{1x}{\sigma}_{4x}+{\sigma}_{1y}{\sigma}_{4y}-{\sigma}_{1z}{\sigma}_{4z})\hat{S}_3\cdot\vec{\sigma}_4\Big)\nonumber\\
&&=\dfrac{1}{4}\times\dfrac{1}{(2S+1)^2}\Big(\mathbb{1}-\frac{\alpha\beta}{S^2}(-S_{2x}S_{3x}+S_{2y}S_{3y}-S_{2z}S_{3z})\Big).
\end{eqnarray}
The factor of $\frac{1}{4}$ represents the probability with which this state is prepared.
\end{enumerate}

\end{appendix}

\begin{thebibliography}{10}

\bibitem{Bell64}
J.~S. Bell.
\newblock On the einstein podolsky rosen paradox.
\newblock {\em Physics}, 1:195, 1964.

\bibitem{Horodecki09}
Ryszard Horodecki, Pawe\l{} Horodecki, Micha\l{} Horodecki, and Karol
  Horodecki.
\newblock Quantum entanglement.
\newblock {\em Rev. Mod. Phys.}, 81:865--942, Jun 2009.

\bibitem{Wiseman07}
H.~M. Wiseman, S.~J. Jones, and A.~C. Doherty.
\newblock Steering, entanglement, nonlocality, and the einstein-podolsky-rosen
  paradox.
\newblock {\em Phys. Rev. Lett.}, 98:140402, Apr 2007.

\bibitem{Ollivier01}
Harold Ollivier and Wojciech~H. Zurek.
\newblock Quantum discord: A measure of the quantumness of correlations.
\newblock {\em Phys. Rev. Lett.}, 88:017901, Dec 2001.

\bibitem{Bennett93}
Charles~H. Bennett, Gilles Brassard, Claude Cr\'epeau, Richard Jozsa, Asher
  Peres, and William~K. Wootters.
\newblock Teleporting an unknown quantum state via dual classical and
  einstein-podolsky-rosen channels.
\newblock {\em Phys. Rev. Lett.}, 70:1895--1899, Mar 1993.

\bibitem{Bennett92}
Charles~H. Bennett and Stephen~J. Wiesner.
\newblock Communication via one- and two-particle operators on
  einstein-podolsky-rosen states.
\newblock {\em Phys. Rev. Lett.}, 69:2881--2884, Nov 1992.

\bibitem{Bennett01}
Charles~H. Bennett, David~P. DiVincenzo, Peter~W. Shor, John~A. Smolin,
  Barbara~M. Terhal, and William~K. Wootters.
\newblock Remote state preparation.
\newblock {\em Phys. Rev. Lett.}, 87:077902, Jul 2001.

\bibitem{Vazirani14}
Umesh Vazirani and Thomas Vidick.
\newblock Fully device-independent quantum key distribution.
\newblock {\em Phys. Rev. Lett.}, 113:140501, Sep 2014.

\bibitem{Zukowski93}
M.~\ifmmode~\dot{Z}\else \.{Z}\fi{}ukowski, A.~Zeilinger, M.~A. Horne, and
  A.~K. Ekert.
\newblock ``event-ready-detectors'' bell experiment via entanglement swapping.
\newblock {\em Phys. Rev. Lett.}, 71:4287--4290, Dec 1993.

\bibitem{Popescu94a}
Sandu Popescu.
\newblock Bell's inequalities versus teleportation: What is nonlocality?
\newblock {\em Phys. Rev. Lett.}, 72:797--799, Feb 1994.

\bibitem{Huang11}
Yun-Feng Huang, Bi-Heng Liu, Liang Peng, Yu-Hu Li, Li~Li, Chuan-Feng Li, and
  Guang-Can Guo.
\newblock Experimental generation of an eight-photon
  greenberger--horne--zeilinger state.
\newblock {\em Nature communications}, 2(1):1--6, 2011.

\bibitem{Monz11}
Thomas Monz, Philipp Schindler, Julio~T. Barreiro, Michael Chwalla, Daniel
  Nigg, William~A. Coish, Maximilian Harlander, Wolfgang H\"ansel, Markus
  Hennrich, and Rainer Blatt.
\newblock 14-qubit entanglement: Creation and coherence.
\newblock {\em Phys. Rev. Lett.}, 106:130506, Mar 2011.

\bibitem{Yao12}
Xing-Can Yao, Tian-Xiong Wang, Ping Xu, He~Lu, Ge-Sheng Pan, Xiao-Hui Bao,
  Cheng-Zhi Peng, Chao-Yang Lu, Yu-Ao Chen, and Jian-Wei Pan.
\newblock Observation of eight-photon entanglement.
\newblock {\em Nature photonics}, 6(4):225--228, 2012.

\bibitem{Dakic12}
Borivoje Daki{\'c}, Yannick~Ole Lipp, Xiaosong Ma, Martin Ringbauer, Sebastian
  Kropatschek, Stefanie Barz, Tomasz Paterek, Vlatko Vedral, Anton Zeilinger,
  {\v{C}}aslav Brukner, et~al.
\newblock Quantum discord as resource for remote state preparation.
\newblock {\em Nature Physics}, 8(9):666--670, 2012.

\bibitem{Madhok13}
Vaibhav Madhok and Animesh Datta.
\newblock Quantum discord as a resource in quantum communication.
\newblock {\em International Journal of Modern Physics B}, 27(01n03):1345041,
  2013.

\bibitem{Fonseca19}
Alejandro Fonseca.
\newblock High-dimensional quantum teleportation under noisy environments.
\newblock {\em Phys. Rev. A}, 100:062311, Dec 2019.

\bibitem{Perez16}
Benjamin Perez-Garcia, Melanie McLaren, Sandeep~K Goyal, Raul~I
  Hernandez-Aranda, Andrew Forbes, and Thomas Konrad.
\newblock Quantum computation with classical light: Implementation of the
  deutsch--jozsa algorithm.
\newblock {\em Physics Letters A}, 380(22-23):1925--1931, 2016.

\bibitem{Garcia18}
Benjamin Perez-Garcia, Raul~I. Hernandez-Aranda, Andrew Forbes, and Thomas
  Konrad.
\newblock The first iteration of grover's algorithm using classical light with
  orbital angular momentum.
\newblock {\em Journal of Modern Optics}, 65(16):1942--1948, 2018.

\bibitem{Spreeuw01}
Robert J.~C. Spreeuw.
\newblock Classical wave-optics analogy of quantum-information processing.
\newblock {\em Phys. Rev. A}, 63:062302, May 2001.

\bibitem{Goyal13}
Sandeep~K. Goyal, Filippus~S. Roux, Andrew Forbes, and Thomas Konrad.
\newblock Implementing quantum walks using orbital angular momentum of
  classical light.
\newblock {\em Phys. Rev. Lett.}, 110:263602, Jun 2013.

\bibitem{Bharath14}
H.~M. Bharath and V.~Ravishankar.
\newblock Classical simulation of entangled states.
\newblock {\em Phys. Rev. A}, 89:062110, Jun 2014.

\bibitem{Adhikary16}
Soumik Adhikary, Ipsit~Kumar Panda, and V.~Ravishankar.
\newblock Super-quantum states in su(2) invariant level systems.
\newblock {\em Annals of Physics}, pages~--, 2016.

\bibitem{Molina2007twisted}
Gabriel Molina-Terriza, Juan~P Torres, and Lluis Torner.
\newblock Twisted photons.
\newblock {\em Nature physics}, 3(5):305--310, 2007.

\bibitem{erhard2018twisted}
Manuel Erhard, Robert Fickler, Mario Krenn, and Anton Zeilinger.
\newblock Twisted photons: new quantum perspectives in high dimensions.
\newblock {\em Light: Science \& Applications}, 7(3):17146--17146, 2018.

\bibitem{Radcliffe71}
J~M Radcliffe.
\newblock Some properties of coherent spin states.
\newblock {\em Journal of Physics A: General Physics}, 4(3):313--323, may 1971.

\bibitem{Werner89}
R.~F. Werner.
\newblock Quantum states with einstein-podolsky-rosen correlations admitting a
  hidden-variable model.
\newblock {\em Physical Review A}, 40(8):4277, October 1989.

\bibitem{Nielsen00}
Michael~A Nielsen and Isaac~L Chuang.
\newblock Quantum information and quantum computation.
\newblock {\em Cambridge: Cambridge University Press}, 2(8):23, 2000.

\bibitem{Hillery96}
V.~Bu\ifmmode~\check{z}\else \v{z}\fi{}ek and M.~Hillery.
\newblock Quantum copying: Beyond the no-cloning theorem.
\newblock {\em Phys. Rev. A}, 54:1844--1852, Sep 1996.

\bibitem{bock2016highly}
Matthias Bock, Andreas Lenhard, Christopher Chunnilall, and Christoph Becher.
\newblock Highly efficient heralded single-photon source for telecom
  wavelengths based on a ppln waveguide.
\newblock {\em Optics express}, 24(21):23992--24001, 2016.

\bibitem{shen2019optical}
Yijie Shen, Xuejiao Wang, Zhenwei Xie, Changjun Min, Xing Fu, Qiang Liu, Mali
  Gong, and Xiaocong Yuan.
\newblock Optical vortices 30 years on: Oam manipulation from topological
  charge to multiple singularities.
\newblock {\em Light: Science \& Applications}, 8(1):1--29, 2019.

\bibitem{padgett2017orbital}
Miles~J Padgett.
\newblock Orbital angular momentum 25 years on.
\newblock {\em Optics express}, 25(10):11265--11274, 2017.

\bibitem{Ma14}
Lingyu Ma and Xiaolong Su.
\newblock Remote transfer of gaussian quantum discord.
\newblock {\em Optics express}, 22(13):15894--15903, 2014.

\bibitem{Xie15}
Chuanmei Xie, Yimin Liu, Hang Xing, Jianlan Chen, and Zhanjun Zhang.
\newblock Quantum correlation swapping.
\newblock {\em Quantum Information Processing}, 14(2):653--679, 2015.

\bibitem{Lutkenhaus99}
N.~L\"utkenhaus, J.~Calsamiglia, and K.-A. Suominen.
\newblock Bell measurements for teleportation.
\newblock {\em Phys. Rev. A}, 59:3295--3300, May 1999.

\bibitem{Ghosh01}
Sibasish Ghosh, Guruprasad Kar, Anirban Roy, Aditi Sen(De), and Ujjwal Sen.
\newblock Distinguishability of bell states.
\newblock {\em Phys. Rev. Lett.}, 87:277902, Dec 2001.

\bibitem{Calsamiglia02}
John Calsamiglia.
\newblock Generalized measurements by linear elements.
\newblock {\em Phys. Rev. A}, 65:030301, Feb 2002.

\bibitem{Cao20}
Cong Cao, Li~Zhang, Yu-Hong Han, Pan-Pan Yin, Ling Fan, Yu-Wen Duan, and
  Ru~Zhang.
\newblock Complete and faithful hyperentangled-bell-state analysis of photon
  systems using a failure-heralded and fidelity-robust quantum gate.
\newblock {\em Optics express}, 28(3):2857--2872, 2020.

\bibitem{Fan21}
Ling Fan and Cong Cao.
\newblock Deterministic cnot gate and complete bell-state analyzer on
  quantum-dot-confined electron spins based on faithful quantum nondemolition
  parity detection.
\newblock {\em JOSA B}, 38(5):1593--1603, 2021.

\end{thebibliography}

\end{document}